\documentclass{article}
\usepackage[utf8]{inputenc}
\usepackage{graphicx}
\graphicspath{ {./images/} }
\usepackage[margin=0.9in]{geometry}
\usepackage{natbib}
\usepackage{amsmath}
\usepackage{amsfonts}
\usepackage{amsthm}
\usepackage{xcolor,colortbl}
\usepackage{multicol}
\usepackage{lscape}
\usepackage{blindtext}
\usepackage{booktabs}
\usepackage{lipsum}
\usepackage{subfigure}
\newcommand{\comment}[1]{}
\usepackage{caption}
\usepackage{subfig}
\usepackage{float}
\usepackage{enumerate}
\usepackage{bm}
\usepackage{mathtools}
\newcolumntype{P}[1]{>{\raggedright\arraybackslash}p{#1}}

\usepackage{algorithm, algorithmicx, algpseudocode}

\usepackage{authblk}

\title{How to Add Baskets to an Ongoing Basket Trial with Information Borrowing}
\author[1]{Libby Daniells}
\author[2]{Pavel Mozgunov}
\author[3]{Alun Bedding}
\author[4]{Helen Barnett}
\author[2,5]{Thomas Jaki}
\affil[1]{STOR-i Centre for Doctoral Training, Department of Mathematics and Statistics, Lancaster University, Lancaster, UK}
\affil[2]{MRC Biostatistics Unit, University of Cambridge, Cambridge, UK}
\affil[3]{Roche Products Ltd, Welwyn Garden City, UK}
\affil[4]{Department of Mathematics and Statistics, Lancaster University, Lancaster, UK}
\affil[5]{University of Regensburg}
\date{}                     
\begin{document}
\maketitle
\begin{abstract}
Basket trials test a single therapeutic treatment on several patient populations under one master protocol. A desirable adaptive design feature in these studies may be the incorporation of new baskets to an ongoing study. Limited basket sample sizes can cause issues in power and precision of treatment effect estimates which could be amplified in added baskets due to the shortened recruitment time. While various Bayesian information borrowing techniques have been introduced to tackle the issue of small sample sizes, the impact of including new baskets in the trial and into the borrowing model has yet to be investigated. We explore approaches for adding baskets to an ongoing trial under information borrowing and highlight when it is beneficial to add a basket compared to running a separate investigation for new baskets. We also propose a novel calibration approach for the decision criteria that is more robust to false decision making. Simulation studies are conducted to assess the performance of approaches which is monitored primarily through type I error control and precision of estimates. Results display a substantial improvement in power for a new basket when information borrowing is utilized, however, this comes with potential inflation of error rates which can be shown to be reduced under the proposed calibration procedure.
\end{abstract}


\section{Introduction}\label{sec:Intro}
In the oncology setting, significant research into cancer genomics and understanding the underlying genetic cause of disease has catapulted the field of personalised medicine, within which treatments are targeted to a specific genetic makeup, to the forefront of clinical trial design \citep{Genomics}. This shift away from disease specific treatments towards genetically targeted treatments has led to the development of basket clinical trials. 

Typically, basket trials are implemented in early stages of the drug development process in order to determine if a treatment is efficacious against each of the individual baskets on the trial \citep{BasketTrial}. They often consist of a single treatment arm using a small number of patients. Basket trials are a form of master protocol in which a single treatment is administered to patients across different disease types, all of whom possess the same genetic aberration. Different disease type sub-populations form their own treatment basket. 

One of the main benefits of basket trials is how they allow for testing of treatments on rare diseases that would not traditionally warrant their own investigation due to their limited sample size \citep{cbhm} and financial and time constraints. By allowing for testing on multiple disease types in a single study, the drug development process is substantially expedited. These basket trials also provide flexibility by utilizing adaptive design features which allow for modification of the design and analysis while the study is still ongoing. Such modifications include interim analysis with futility and efficacy stopping, sample size adjustment, or as is the focus of this work, the addition of a single or multiple baskets to an ongoing trial. This situation arises when it is identified that a new group of patients may benefit from the treatment, where these patients harbour the genetic aberration under investigation but suffer from a different disease type. 

Several prominent clinical trials have utilized the addition of a treatment arm/basket. The VE-BASKET trial \citep{VEExample}, exploring the effect of Vemurafeib on cancers with the BRAFV600 mutation, is an example of such a study in which two new baskets were formed of patients on the trial due to sufficient accrual rates of patients of two new cancer types. Likewise, the basket trial looking at the treatment of tucatinib and trastuzab on solid tumors with the HER2 alteration \citep{HER2Example} allowed the opening of disease-specific cohorts during the trial as a part of the trial protocol.  These trial examples act as motivation of the work presented in this paper, with the purpose to explore methodology for handling such additions of treatment groups.

Although basket trials are desirable as they allow the testing of rare diseases, this does introduce challenges in cases where sample sizes are limited. In such cases issues such as lack of statistical power and precision of estimates arise. This can be amplified in baskets that are added part-way through an ongoing trial if the disease type is also rare. The combination of reduced recruitment rate and shorter recruitment time due to the late addition to the trial can result in further reduction in sample sizes compared to baskets that opened at the beginning of the trial.

To tackle the issue of small sample sizes within baskets, Bayesian information borrowing methods were proposed to be used for basket trials. These methods utilize the assumption that, as patients across baskets share the same genetic mutation, they will have a similar response to the treatment. As such, patients are `exchangeable' between baskets, meaning patients can be moved between treatment baskets without changing the overall treatment effect estimates \citep{exchangeability}. One can use this assumption to draw on information from one basket when making inference in another, thus potentially improving power and precision of estimates, particularly in the presence of small sample sizes. 

Over recent years, several prominent methods for information borrowing in basket trials have been proposed. These include the Bayesian hierarchical model \citep[BHM,][]{Hierarchical} and several adaptations to this method, such as the calibrated Bayesian hierarchical model \citep[CBHM,][]{cbhm} which defines the prior on the borrowing parameter as a function of homogeneity, the exchangeability-nonexchangeability model \citep[EXNEX,][]{exnex} which allows for flexible borrowing between subsets of baskets and the modified exchangeability-nonexchangeability model \citep[mEXNEX$_c$,][]{mEXNEXc} which modifies the EXNEX model to account for homogeneity/heterogeneity between baskets. 

This paper proposes and investigates different approaches for the analysis of newly added baskets under an information borrowing structure, which primarily utilizes the EXNEX model. Although one approach for adding baskets could be to analyse new baskets akin to baskets that begun the trial, resulting in a problem mathematically equivalent to a case of unequal sample sizes, this paper explores additional approaches. The motivation for the considered approaches is the belief that additional baskets could negatively impact the type I error rate and other operating characteristics of existing baskets should results be heterogeneous, however, potential power can be gained by utilizing information borrowing when analyzing any new baskets in cases of homogeneity. In addition, it is important to minimise error inflation in new baskets too. Results of thorough simulation studies are provided in this paper to compare approaches for adding baskets in order to identify when and how it can be beneficial to add a new basket to an ongoing trial as opposed to running a separate investigation for the new basket(s). Results support the rational of treating this problem as separate to simply a case of unequal sample sizes, as although there is no one approach that is more beneficial across all scenarios, it is clear that an approach that analyses a new basket as independent whilst retaining information borrowing between existing baskets has better error control when there is heterogeneity between new and existing baskets, however, it is also clear that significant power can be gained via borrowing between all baskets when the new basket is homogeneous to existing ones. We also show that this gain in power in the new basket is related to the number of existing baskets that are effective to treatment.  

When implementing Bayesian information borrowing methods, once the model is fit to the data, posterior probabilities are computed and compared to some pre-defined cut-off value in order to determine whether or not a treatment is efficacious in each of the baskets on the trial. Traditionally, these cut-off values are calibrated through simulation studies under a global null scenario, where all baskets have a truly ineffective response rate, in order to control the basket specific type I error rate to a nominal level. However, when this cut-off value is then applied to cases where at least one basket is non-null, it is not guaranteed that error rates will remain controlled at the nominal level when information borrowing is utilized \citep{KOPP}. In fact, inflation in error rates often occurs in cases of heterogeneity as borrowing information causes shifts in the posterior probabilities away from the true treatment effect, resulting in more errors. This brings into question whether calibrating under the global null is sufficient, as more often than not, there is an expectation that the treatment is efficacious in at least one basket. We propose a novel calibration technique, called the \textbf{R}obust \textbf{Ca}libration \textbf{P}rocedure (RCaP), which controls the type I error rate \textit{on average} across several possible true response rate data scenarios, with the potential to weight scenarios based on importance and prior likelihood of occurring in the trial. Presented in this paper is a comparison between simulation studies that calibrate using the traditional approach and using this novel procedure. Noticeable benefits of RCaP are observed with a substantial reduction in error rates alongside only a small loss in power relative to the targeted value in a small handful of cases. 

This paper is structured as follows, we begin with introducing a motivating example, the VE-BASKET study, in which a basket design was implemented with the addition of two new baskets. We then describe analysis models, approaches for the analysis of newly added baskets, and outline the novel calibration procedure, RCaP. Results of several simulation are presented starting with a comparison of calibration techniques, followed by results of simulation studies to compare performance of approaches for the addition of a newly identified basket. Finally, we explore the effect of timing of addition of the new basket on the performance of approaches, again through simulation studies. 

\subsection{Motivating Trial: The VE-BASKET Study}\label{sec:MotivatingExample}

\vspace{-1em}
\begin{figure}[ht]
    \centering 
    \includegraphics[width=0.8\textwidth]{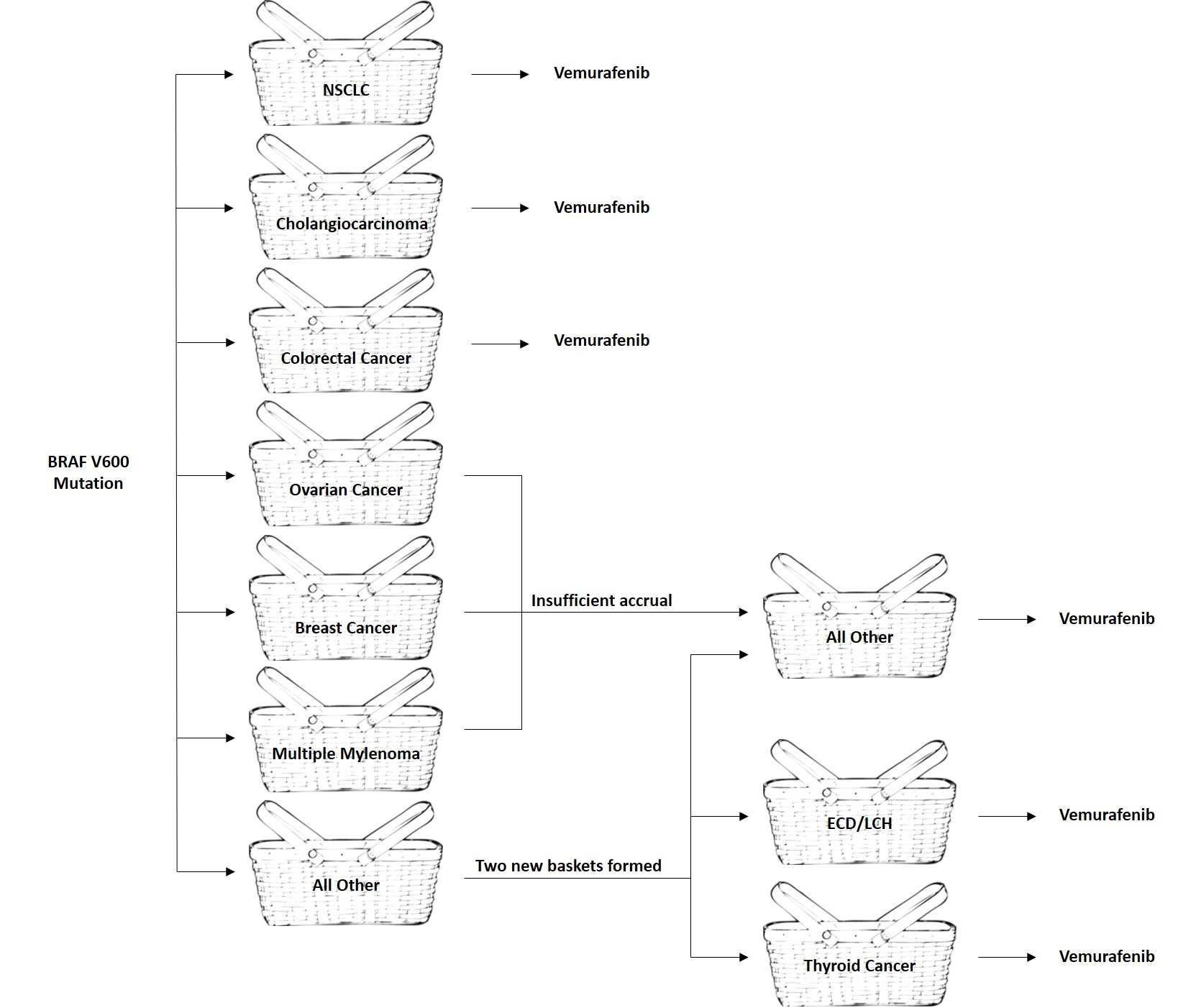}
    \caption{VE-BASKET Trial Design}
    \label{fig:VEBASKET}
\end{figure}

The VE-BASKET trial was a phase II basket trial, investigating the effect of Vemurafeib on multiple cancer types possessing the BRAFV600 mutation \citep{VEExample}. A total of 122 patients were enrolled across all baskets, with efficacy evaluated after 8 weeks of treatment. The primary endpoint in this study was the overall response rate (ORR) with a null response rate of 15\% indicating inactivity and target response rate of 45\%. A response rate of 35\% was considered low but still indicative of a response. Sample sizes of 13 patients per basket were obtained through a Simon's two stage design \citep{Simon} based on 80\% power and 10\% type I error rate.

The trial opened with 6 disease specific baskets: non-small-cell lung cancer (NSCLC), ovarian cancer, colorectal cancer, cholangiocarcinoma, breast cancer and multiple myeloma. Also present was an `all other' basket consisting of patients with different disease types with the BRAFV600 mutation. This initial trial structure was adapted based on recruitment rates, with the breast cancer, ovarian cancer and multiple myeloma baskets closing due to insufficient accrual, where patients were then moved to the `all other' basket for analysis. However, due to sufficient number of patients in the `all other' group, two new baskets were formed and added to the trial: an Edrheim-Chester disease or Langerhans'-cell histiocytosis (ECD/LCH) basket and a anaplastic thyroid cancer basket. Figure \ref{fig:VEBASKET} displays the general trial schematic.

Although not newly identified baskets, with baskets forming from patients already on the trial, this highlights the flexible nature one would like in a basket trial in that baskets can be added in the ongoing study, hence bringing about the question on how to analyse such baskets.

\section{Methodology}\label{sec:Methods}
\subsection{Setting}
This paper focuses on a single treatment arm within each basket and considers binary endpoints, in which a patient either responds positively to a treatment or does not. Consider a basket trial with a total of $K$ baskets. Denote the responses in basket $k$ by $Y_k$ which follows a binomial distribution, $Y_k\sim\text{Binomial}(n_k,p_k)$, with $n_k$ and $p_k$ indicating the sample size and response rate in basket $k$ respectively. Interest lies in estimating the unknown response rate $p_k$. Denote $q_0$ and $q_1$ as the null and target response rate respectively.

Now consider a case where baskets of patients are added to an ongoing trial and thus split the $K$ baskets into two sets. Let $K_0$ be the total number of existing baskets that began the trial, thus having $K^{\prime}=K-K_0$ new baskets added part way through the study. Existing baskets are indexed $k_0=1,\ldots,K_0$ and new baskets $k^{\prime}=K_0+1,\ldots,K$. Note that a new basket, $k^{\prime}$, may be added at any time during the study and it is not required that all new baskets be added at the same time.  

The objective is to test the family of hypotheses:
\begin{align*}
    &H_0\;:\;p_{k_0}\leq q_0\hspace{3em}vs.\hspace{3em}H_a\;:\;p_{k_0}>q_0,&k_0&=1,\ldots,K_0,\\
    &H_0\;:\;p_{k^{\prime}}\leq q_0\hspace{3em}vs.\hspace{3em}H_a\;:\;p_{k^{\prime}}>q_0,&k^{\prime}&=K_0+1,\ldots,K.
\end{align*}

To test these hypotheses, a Bayesian framework is utilized. In the trial protocol prior distributions are defined for parameters of interest to reflect the initial beliefs of clinicians. Posterior distributions are then produced by updating the prior distributions with the observed response data, $D$, collected at the conclusion of the trial. Posterior probabilities are used to determine the efficacy of the treatment on each of the individual baskets in the trial, as such, the treatment is deemed effective in an existing basket $k_0$ if $\mathbb{P}(p_{k_0}>q_0|D)>\Delta_{k_0}$ and effective in a new basket $k^{\prime}$ if $\mathbb{P}(p_{k^\prime}>q_0|D)>\Delta_{k^{\prime}}$. Both cut-off values $\Delta_{k_0}$ and $\Delta_{k^{\prime}}$ are typically determined through \textit{calibration} in order to control some metric, often related to false decision making, at a nominal level. Traditionally this calibration is done under a global null scenario in which all baskets are ineffective to treatment, in order to control the basket-specific type I error rate to a nominal level.

\subsection{Models}\label{Sec:Models}
Borrowing models utilize the exchangeability assumption, that as patients across all baskets share a common genetic component, their response to treatment will be similar, thus information can be shared between all baskets. The Bayesian hierarchical model (BHM) first outlined by Berry et al\cite{Hierarchical} is a key basis for many borrowing models, utilizing this full exchangeability assumption to borrow between all baskets in the trial. The BHM is specified such that the log-odds of the response rate for each basket follows a normal distribution centered around a common mean $\mu$ with variance $\sigma^2$. Hyper-priors are then placed on the parameters $\mu$ and $\sigma^2$.
\begin{align}\label{model:BHM}
    &Y_k\;\sim\;\text{Binomial}(n_k,p_k),\quad\quad k=1,\ldots,K\nonumber\\
    &\theta_k\;=\;\text{logit}(p_k)\;\sim\;\text{N}(\mu,\sigma^2),\nonumber\\
    &\mu\;\sim\;\text{N}(\text{logit}(q_0),\nu_\mu),\quad\sigma\sim\text{g}(\cdot).
\end{align}
 Under this BHM, borrowing occurs between all baskets and thus estimates of response rates are shrunk towards the common mean $\mu$ with the degree of shrinkage controlled by $\sigma^2$. As $\sigma^2$ tends to 0, borrowing moves towards complete pooling of results, however, as it tends to infinity a stratified analysis is conducted on each basket. 

Issues arise in these Bayesian hierarchical models when the exchangeability assumption is violated, which occurs in the presence of heterogeneous baskets. In such cases, under models that borrow information between all baskets, the type I error rate is likely to inflate as the posterior probabilities are pulled towards the common mean and away from the true treatment effect. 

In the literature there are several prominent information borrowing models utilizing Bayesian methodology in the basket trial setting such as the calibrated Bayesian hierarchical model \citep[CBHM,][]{cbhm}, exchangeability-nonexchangeability model \citep[EXNEX,][]{exnex}, modified exchangeability-nonexchangeability model \citep[mEXNEX$_c$,][]{mEXNEXc}, MUCE design \citep{MUCE}, the RoBoT design \citep{RoBot}, Liu's two-path approach \citep{Liu2} and a Bayesian model averaging approach \citep[BMA,][]{BMA}. This paper focuses primarily on the EXNEX model due to it's beneficial incorporation of a non-exchangeability component to the standard hierarchical model outlined in (\ref{model:BHM}), within which baskets are analysed independently. This allows borrowing between just a subset of baskets, reducing the impact of heterogeneous baskets on the type I error rate.

\subsubsection{Exchangeability-Nonexchangeability Model}\label{sec:exnex}
The exchangeability-nonexchangeability (EXNEX) model, proposed by Neuenschwander et al\cite{exnex}, is an extension to the standard BHM that allows for more flexibility in borrowing between baskets. The model has two components:
\begin{enumerate}
    \item EX (exchangeable component): with prior probability $\pi_k$, basket $k$ is exchangeable and a Bayesian hierarchical model is applied. Information borrowing is therefore conducted between all baskets assigned to the exchangeable component. 
    \item NEX (nonexchangeable component): with prior probability $1-\pi_k$, basket $k$ is nonexchangeable with any other basket, and as a result is analysed independently.
\end{enumerate}

\begin{center}\vspace{-2em}
\begin{minipage}[t]{.45\textwidth}
  \begin{align*}
         &Y_{k}\sim\text{Binomial}(n_k,p_k), \quad\quad k=1,\ldots,K\nonumber\\
    &\theta_k=\log\Big(\frac{p_k}{1-p_k}\Big),\\
    &\theta_k=\delta_kM_{1k}+(1-\delta_k)M_{2k},\\
    &\delta_k\sim\text{Bernoulli}(\pi_k),
  \end{align*}
\end{minipage}%
\begin{minipage}[t]{0.45\textwidth}
\begin{align}\label{eq:EXNEX}
    &M_{1k}\sim\text{N}(\mu,\sigma^2),\hspace{1em}\text{(EX)}\nonumber\\
      &\mu\sim\text{N}(\text{logit}(q_0),\nu_\mu),\nonumber\\
    &\sigma\sim g(\cdot),\nonumber\\
     &M_{2k}\sim \text{N}(m_k,\nu_k).\hspace{1em}\text{(NEX)}
\end{align}
    \end{minipage}
    \end{center}

\noindent As outlined in the model specification (\ref{eq:EXNEX}), the EX component has the form of a Bayesian hierarchical model with the log-odds of the response rates for each basket following a normal distribution, centred around a common mean $\mu$ with variance $\sigma^2$. The prior on $\mu$ is centred around the average null response rate across the baskets with a large variance, whilst the prior on $\sigma$, $g(\cdot)$, is more widely debated with Inverse-Gamma, Half-Normal or Half-Cauchy priors implemented across the literature. It is suggested that a Half-Normal(0,1) prior is to be placed on $\sigma$ as this allows for anywhere between a small and very large amount of heterogeneity between baskets\cite{exnex}. 

As a BHM is implemented, estimates of the response rates for each basket in the EX component are shrunk towards the common mean $\mu$, with shrinkage controlled by $\sigma^2$, leaving the potential for increased error rates in cases where the exchangeability assumption is broken. However, the effect of this is reduced by the incorporation of the nonexchangeability component, which allows baskets to be analyzed independently. 

For the NEX component, prior parameters on the logit transformed response rates are basket specific, and thus each basket is analysed independent of one another. Neuenschwander et al\cite{exnex} propose setting the parameters as follows:
\begin{equation}\label{eq:exnexnexpar}m_k=\log\bigg(\frac{\rho_k}{1-\rho_k}\bigg),\;\;\;\nu_k=\frac{1}{\rho_k}+\frac{1}{1-\rho_k},\end{equation} where $\rho_k$ is a plausible guess for the true response rate in basket $k$.

The prior probabilities, $\pi_k$, for assignment to the EX/NEX component are selected prior to the trial. There is often little to no information available on the probability of exchangeability of baskets before the trial, so it is suggested to fix $\pi_k=0.5$ for all $k$ baskets. Alternatively, a Dirichlet prior could be placed on these values, however, Neuenschwander et al\cite{neuenschwander2023fixed} prove that only the mean of the weight distribution affects inference in this case.

\subsubsection{Independent Analysis}\label{sec:independent}
Alongside borrowing models, an independent stratified analysis is also implemented in this paper. Under an independent model, no information is shared between baskets and each basket is analysed independently of one another. Assuming a total of $K$ baskets:
\begin{align}\label{eq:ind}
    &Y_k\;\sim\;\text{Binomial}(n_k,p_k), \quad\quad k=1,\ldots,K\nonumber\\
    &\theta_k\;=\;\text{logit}(p_k),\nonumber\\
    &\theta_k\;\sim\;\text{N}(\text{logit}(q_{0k}),\sigma_k^2),
\end{align}
where $q_{0k}$ denotes the null response rate in basket $k$. The logit transformation of the response rates is taken to avoid boundary issues when $p_k$ is close to 0 or 1 and a slightly informative normal prior is placed on this transformed parameter, with mean based on the null response rate but with a large variance, $\sigma^2_k$ \citep{mEXNEXc}.

Unlike Bayesian information borrowing models, no inflation in error rates are expected as under stratified analysis, no pull towards a common mean is present because heterogeneity between baskets is irrelevant. Thus one would expect error rates to remain at the nominal level that the design parameters have been calibrated to achieve. 

\subsection{Approaches for Adding A Basket}\label{sec:Approaches}
We now propose four different approaches for the calibration and analysis of newly added baskets to an ongoing basket trial. In all four cases existing baskets are analyzed through an EXNEX model, however, treatment of the new basket varies. Approaches are outlined below, as well as, summarized in Table \ref{tab:approachesforaddingcondensed}.
\begin{enumerate}
    \item \textbf{IND} - INDependent analysis of the new basket. 
    
    Calibrate $\Delta_{k_0}$ based on an EXNEX (as in Equation (\ref{eq:EXNEX})) model applied to the $K_0$ existing baskets and $\Delta_{k^{\prime}}$ based on an independent model applied to each of the $K^{\prime}$ new baskets (as in Equation (\ref{eq:ind})). Analyse results in the same way.

    Analysing the new basket as independent may be considered desirable as it eliminates potential negative effects of smaller sample sizes in new baskets on inference in existing baskets.
    
    \item\textbf{UNPL} - UNPLanned addition of a new basket.
    
    Calibrate $\Delta_{k_0}$ based on an EXNEX model applied to the $K_0$ existing baskets. When conducting analysis borrow between all $K$ baskets through an EXNEX model and for cases of equal sample size set $\Delta_{k^{\prime}}=\Delta_{k_0}$ for the new basket. If the sample size is unequal for the $K'$ new baskets compared to the $K$ existing baskets, set $\Delta_{k^{\prime}}=\Delta_{i_0}$ where existing basket $i$ has sample size $n_i$ closest to the sample size of the new basket $k^{\prime}$, i.e. $i=\arg\min_i\{|n_i-n_{k^{\prime}}|\}$. 
    
    This is a naive analysis in that cut-off values are not adjusted despite the additional baskets. This may occur when an addition is not planned for but once it occurs it is believed that borrowing information between all baskets, new and old, can improve inference.
    
    \item \textbf{PL1} - PLanned addition of a new basket in which a single EXNEX model is applied. 
    
    Calibrate $\Delta_{k_0}$ and $\Delta_{k^{\prime}}$ assuming that new baskets will be added during the study. To calibrate and analyze, borrow between all $K$ baskets new and existing through an EXNEX model. Effectively, this equates to calibration under unequal sample sizes and has two subsets:
    \begin{enumerate}
        \item[(a)] The time of addition of the new basket(s) is known. In this case, the sample sizes, $n_k$, for each of the $k=1,\ldots,K$ baskets are known and fixed in the calibration procedure. 
        \item[(b)] The time of addition of the new basket(s) is unknown. In this case further simulation studies are required to explore the effect of sample size on operating characteristics. It may be desirable to calibrate based on the least favourable configuration to ensure better error control.
    \end{enumerate}
    
    The situation in which it is known for certain that new baskets will be added may occur if it is apparent that a basket of patients will benefit from the study, however, is not ready in time for the commencement of the trial, whether this is due to logistical issues or diagnostic techniques or some other factor. Thus it is planned to add the basket at a later time. This time of addition may be fixed as in PL1(a) but it may be desirable to add a basket as soon as it is available, thus falling into the case of PL1(b) when timing of addition is unknown. 
    
    \item\textbf{PL2} - PLanned addition of a new basket in which two EXNEX models are applied.
    
    Calibrate $\Delta_{k_0}$ based on an EXNEX model applied to just the $K_0$ existing baskets and thus, when analysing the existing baskets, do not borrow from any new baskets. Calibrate $\Delta_{k^{\prime}}$ based on an EXNEX model applied to all $K$ baskets, so under analysis for new baskets borrow information from all baskets. This results in two EXNEX models and has two subsets:
    \begin{enumerate}
        \item[(a)] The time of addition of the new basket(s) is known. Again, sample sizes $n_k$ for all $k=1,\ldots,K$ baskets are known and thus fixed in the calibration procedure. 
        \item[(b)] The time of addition of the new basket(s) is unknown. As in PL1(b), further simulation studies would be required to explore the effect of $n_{k^{\prime}}$ on operating characteristics and adjust calibration accordingly.
    \end{enumerate}
    
    As in IND, analysing baskets in this way will eliminate the effect of reduced sample sizes in new baskets on estimation of response rates in existing baskets. However, by allowing full information borrowing between all baskets when analysing the new baskets, one may combat the issue of lack of statistical power and precision of estimates that arises due to the limited sample size.  
\end{enumerate}
\vspace{0em}
\begin{table}[ht]
\centering
\caption{Summary of approaches for analysis and calibration when adding a basket where $k_0$ denotes existing baskets and $k^{\prime}$ denotes new baskets.}\label{tab:approachesforaddingcondensed}
\begin{tabular}{|l|ll|ll|}
\hline
\bf{Approach}& \multicolumn{2}{c|}{\bf{Calibration of}}& \multicolumn{2}{c|}{\bf{Analysis}}\\
& \multicolumn{1}{c}{$\Delta_{k_0}$} & \multicolumn{1}{c|}{$\Delta_{k^{\prime}}$} & \multicolumn{1}{c}{Existing Baskets} & \multicolumn{1}{c|}{New Baskets}\\
\hline
IND & EXNEX on all $k_0$ & Independent on all $k^{\prime}$ & EXNEX on all $k_0$& Independent on all $k^{\prime}$\\
UNPL & EXNEX on all $k_0$ & $\Delta_{k_0}=\Delta_{k^{\prime}}$ & \multicolumn{2}{c|}{EXNEX on all $k$}\\
PL1 & \multicolumn{2}{c|}{EXNEX on all $k$}& \multicolumn{2}{c|}{EXNEX on all $k$}\\
PL2 & EXNEX on all $k_0$ & EXNEX on all $k$& EXNEX on all $k_0$& EXNEX on all $k$\\
\hline
\end{tabular}
\end{table}%

With the approaches specified in this way, for existing baskets, IND and PL2 are equivalent, as to analyse these baskets under both approaches is to borrow between all $K_0$ existing baskets, independent of any new baskets. Similarly, for new baskets, PL1 and PL2 are equivalent as under both, when calibrating and analysing any of the $K^{\prime}$ new baskets, information is borrowed between all $K$ baskets in the trial through fitting an EXNEX model. Full model specifications and a further table summary are provided in the supplementary material.

\subsection{RCaP: Robust Calibration Procedure}\label{sec:calibrateacross}
A treatment is deemed effective in basket $k$ if the posterior probability that the response rate, $p_k$, is greater than the null, exceeds a cut-off value $\Delta_k$. In a few basket trial cases, such as the work by Zheng and Wason\cite{fixeddelta1} and Ouma et al\cite{fixeddelta2}, these $\Delta_k$ values are fixed at some value, i.e. 0.975, however, an alternative is to calibrate the cut-off value in order to control some operating characteristic to a desirable level. 

For example, Kaizer et al\cite{caldelta1}, Hobbs and Landin\cite{caldelta2}, Chu and Yuan\cite{cbhm}, Jin et al\cite{OC} and Berry et al\cite{Hierarchical}, all followed a conventional approach where $\Delta_k$ was calibrated under a single simulation scenario (i.e. a vector of probabilities that reflect response rates in each of the baskets in the trial), typically this is the global null scenario in which the treatment is ineffective across all baskets. In each of these cases $\Delta_k$ was calibrated to achieve an $100\alpha\%$ type I error rate in each basket under the global null. Despite this, the calibration does not guarantee error control across other scenarios when information borrowing is implemented. When borrowing information from heterogeneous and effective baskets, the posterior probabilities are pulled upwards for baskets with an ineffective response rate, thus increasing the probability of exceeding the calibrated value, $\Delta_k$, compared to under the global null scenario. Therefore, error control is only guaranteed under the global scenario in which $\Delta_k$ was calibrated under, with other scenarios likely to demonstrate undesirable error rate inflation. This is observed in the simulation study conducted by Daniells et al\cite{mEXNEXc}, with the EXNEX model, in the worst case, producing a relative increase in type I error rate of 72.5\% compared to the nominal 10\% level. Similar findings are presented by Jin et al\cite{OC} (236\% relative increase) and Chen and Hsiao\cite{OC2} (135\% relative increase). Although $\Delta_{k}$ is typically calibrated to control the type I error rate, the calibration procedure remains the same for the control of any metric obtained from the posterior density. 

We propose a novel calibration procedure, the Robust Calibration Procedure (RCaP), where as opposed to calibrating under a single global null scenario, $\Delta_k$ is calibrated across numerous potential scenarios so that some metric, $Q$, is controlled \textit{on average} across potential trial outcomes and true response rate vectors. Simulation scenarios may be weighted in importance by their probability of actually occurring in the trial, or if no information is provided on potential outcomes, scenarios can be equally weighted.

Consider a case with $K$ total baskets with $\bm{n}=(n_1,\ldots,n_K)$ being the vector of sample sizes for all baskets $k=1,\ldots,K$. Let there be $M$ simulation scenarios $\bm{p_1},\ldots,\bm{p}_M$ one wishes to calibrate across. Denote true response rates for basket $k$ under scenario $m$ as $p_{mk}$ with $k=1,\ldots,K$ and $m=1,\ldots,M$. The simulation scenarios are vectors consisting of these true response rate probabilities for each of the $K$ baskets, i.e. $\bm{p}_m=(p_{m1},\ldots,p_{mK})$ for all $m=1,\ldots,M$. These scenarios are used in each of the simulation runs for the calibration alongside the basket sample sizes in order to generate data $\bm{X}$ from $\bm{X}\sim F(\bm{p_m},\bm{n})$.

Each of these simulation scenarios may be weighted to reflect importance or likelihood of them actually occurring in the trial. Thus define weights $\omega_m\in\mathbb{N}$ for each scenario $m=1,\ldots,M$. If no weight is required, set $\omega_m=1$ for all $m=1,\ldots,M$. These weights are integer values, with larger values increasing the number of posterior probabilities that contribute to the calibration process for that scenario, thus increasing the scenarios importance in the calibration. Therefore, weights can be interpreted as a ratio across scenarios, and it is this ratio of weights that puts more emphasis on certain simulation scenarios. If required, these weights can be normalized to sum to 1, i.e. $\omega_m/(\sum^M_{i=1}\omega_i)$, however, integer values must be used in Algorithm \ref{alg:cal}.

The generalized novel Robust calibration procedure, taking into account the calibration of any metric, is described in Algorithm \ref{alg:cal}. The algorithm requires the specification of sample sizes and definition of all $M$ simulation scenarios under consideration, alongside their weights of importance, $\omega_m$ for $m=1,\ldots,M$. For each simulation scenario, $\bm{p_m}$, a total of $R$ data sets are generated from $F(\bm{p_m},\bm{n})$. A model is then fit to each of these $R$ data sets to obtain posterior densities. A quantity, $Q$, required to compute the metric of interest is then obtained from these posterior distributions. When the metric of interest is the type I error rate, $Q$ would be the posterior probability of the response rate exceeding the null $\mathbb{P}(p_{mk}>q_0|\bm{X})$ (for basket $k$ under scenario $m$ with $k=1,\ldots,K$ and $m=1,\ldots,M$). To conduct the calibration to control $Q$, the metric under each simulation scenario for each of the $R$ data sets must be stored. In Algorithm \ref{alg:cal}, if a basket $k$ satisfies the basket-specific condition, $T(\cdot)$, these $Q$ values are stored in vectors $\bm{PP_1},\ldots,\bm{P_K}$. These basket-specific conditions may require that the true response rate $p_{mk}$ is null if computing the type I error or non-null if computing the power for example with $T(\cdot)$ being a binary outcome taking value 1 when the criterion is satisfied and 0 otherwise. Weights are introduced in this step as a total of $\omega_m$ copies of $Q$ are stored in each vector $\bm{PP_k}$ for scenario $m=1,\ldots,M$ to define their contribution to the calibration. Finally, to compute cut-off values, $\Delta_k$, the appropriate quantile is taken within each $\bm{PP_k}$ for all $k=1,\ldots,K$. As such, the chosen cut-off value will be the quantile of the combined metrics across all $M$ scenarios under consideration which satisfy the basket-specific criterion (weighted by importance through $\omega_m$), thus controlling the metric $Q$ across all scenarios combined. 

The basket specific criterion, $T(\cdot)$ on each basket $k=1,\ldots,K$ is of importance as it is only simulation scenarios that satisfy this criterion that contribute to the calibration procedure for the cut-off value in basket $k$. For instance, consider the case where the metric of interest is the probability of a type I error, $Q$, is the probability of a type I error, the probability of a type I error can only be computed in cases where the basket $k$ under scenario $m$ has a null response rate, if non-null a type I error cannot occur. Therefore, the basket specific criterion that $p_{mk}$ must be null, i.e. $T(p_{mk}\leq q_0)$, will need to be satisfied for the simulation scenario to contribute to $\bm{PP_k}$ and thus to be incorporated into the calibration procedure. When calibrating for type I error control, as is the focus in this paper, it is therefore important to require that in each of the $M$ simulation scenarios, at least one basket has a null response rate to satisfy the basket specific criterion. The full algorithm applied to control the type I error rate is provided in the supplementary material.

\begin{algorithm}[ht]
\begin{algorithmic}
\caption{RCaP - Calibrate $\Delta_k$ across several simulation scenarios for any metric, $Q$.}\label{alg:cal}
\State \textbf{Data:} Total number of simulation scenarios, $M$, scenarios $\bm{p_1},\ldots,\bm{p_M}$, basket sample sizes $\bm{n}$, number of simulation runs for each scenario, $R$, and integer weights for the scenarios, $\omega_1,\ldots,\omega_M$;
\State \textbf{Initialization:} $\bm{PP_1},\ldots,\bm{PP_K}$ empty vectors\;
\For {$m = 1$ to $M$}
\For {$r=1$ to $R$}
\State Generate data $\bm{X}\sim F(\bm{p_m},\bm{n})$\;
\State Fit information borrowing model to obtain posterior densities\;
\State Compute a quantity, $Q$, obtained from the posterior required for the metric of interest\;
\For {$k=1$ to $K$}
\If {Basket $k$ satisfies the basket specific criterion, $T(\cdot)$}
\For {$j=1$ to $\omega_m$}
\State $\bm{PP_k}=\bm{PP_k}\cup Q$\;
\EndFor
\EndIf
\EndFor
\EndFor
\EndFor
\State $\Delta_k$ = $100(1-\alpha)\%$ quantile of $\bm{PP}_k$ for each basket $k$.\;
\State \Return {Cut-off values $\Delta_k$ for each basket $k$;}
\end{algorithmic}
\end{algorithm}

In cases where sample sizes are equal across a number of baskets, the cut-off values $\Delta_k$ for all baskets of the same size are expected to be equal. To calibrate in this case, define the set of baskets with equal sample size as $E$. Find the basket in $E$ which maximises the number of times the basket-specific criterion $T(\cdot)$ is satisfied across the $M$ simulation scenarios, denote this basket as basket $\epsilon$. Then set $\Delta_k=\Delta_\epsilon$ for all $k\in E$. With the procedure outlined in this way, the RCaP will maximise the number of simulation scenarios that contribute to the calibration. In a special case where \textit{all} baskets are of the same size, for simplicity fix the response rate for the first basket as satisfying the condition $T(\cdot)$ in all $M$ data scenarios, varying the response rate in other baskets. Then set the cut-off values for all baskets as $\Delta_1$ obtained through Algorithm \ref{alg:cal}. This is to ensure that $\Delta_k$ is calibrated across the greatest number of simulation scenarios in order to achieve control of metric $Q$ on average. 

Under this RCaP in order to control for a type I error, one would expect superior error control across non-null cases compared to the traditional calibration under the global null, as the $\Delta_k$ values obtained will likely be more conservative to ensure error control across multiple scenarios. With the increased conservative nature, a decrease in power is also likely.


\section{Simulation Study - A Comparison Between Approaches}\label{sec:SimComparison}
\subsection{General Setting}
In order to explore and compare operating characteristics of the proposed approaches for handling the addition of a new basket to an ongoing trial, numerous simulation studies have been conducted. Simulation studies are split into two categories with the first exploring the case in which scenarios/truth vectors are fixed within the simulation to pre-defined values and secondly, truth vectors are randomly generated within simulation runs. Two lines of comparison are then made within these simulation studies: the typical approach of calibrating under the global null is compared to the RCaP followed by a comparison between the approaches for adding a basket to an ongoing trial. In both of these simulations only subset (a) of PL1 and PL2 in which time of addition is known are considered. However, an exploration into the effect of timing of addition is later provided to assess the performance of PL1(b) and PL2(b). Other simulation studies are provided in the supplementary material including the effect of including a different number of scenarios in the RCaP as opposed to just the partial and global nulls, as well as, a simulation considering a different trial configuration.

We consider the following trial setting. There are $K_0=4$ existing baskets with $K^{\prime}=1$ new basket added part-way through the study. Let the null and target response rates be $q_0=0.2$ and $q_1=0.4$ respectively with a response rate of $q_2=0.3$ indicating a marginally effective response. For existing baskets, sample sizes were fixed at $n_{k_0}=24$ for $k_0=1,2,3$ and 4. For the new basket, as the timing of addition is for now assumed as known, assume equal accrual rates across all $K$ baskets and new and existing and set the sample size as $n_{k^{\prime}}=14$ for $k^{\prime}=5$. This is obtained by guidance of a Simon two-stage design \citep{Simon} with a nominal targeted type I error rate and power of 10\% and 80\% respectively. 

The following operating characteristics are computed within each simulation study:
\begin{itemize}
    \item $\%$ Reject: the percentage of simulated data sets in which the null hypothesis is rejected in each basket. If the true response rate is null then this value is the type I error rate, else it is the power. 
    \item Family-Wise Error Rate (FWER): the percentage of simulated data sets in which the null hypothesis was rejected for at least one truly ineffective basket. 
    \item $\%$ All Correct: the percentage of simulated data sets in which the correct conclusion regarding accepting/rejecting the null was made across all $K$ baskets existing and new. 
    \item Mean Point Estimate: the mean estimate of $p_k$ across the simulated data sets for each of the $K$ baskets presented alongside the standard deviation of these estimates. 
\end{itemize}

All simulations are conducted using the `rjags' package v 4.13 \citep{jags} within RStudio v 1.1.453 \citep{RStudio}. With simulations consisting of 10,000 simulation runs for each data scenario and approach considered. Results presented here focus on the type I error rate and power, however, results of other metrics are included in the supplementary material.

\subsection{Prior Specification}
Throughout the simulations the independent model is specified such that the prior placed on the logit transformation of the response rate $p_k$ follows a Normal distribution: $\theta_k\sim\text{N}(\text{logit}(0.2),10^2)$ and is therefore centred around the null response rate with a large variance. The same prior is placed on $\mu$ in the exchangeability component of the EXNEX model with a $\text{Half-Normal}(0,1)$ prior placed on $\sigma^2$. For the non-exchangeability component, as suggested in Equation (\ref{eq:exnexnexpar}), based on a plausible guess $\rho_k=0.3$ for the true response rate across all baskets $1,\ldots,K$, the prior on the NEX component is a normal distribution with mean -0.85 and variance 4.76. The prior probabilities for assignment to the EX/NEX component are fixed at $\pi_k=0.5$ for all baskets new and existing. Full model specifications are provided in the supplementary material.

\subsection{Fixed Data Scenarios Simulation Setting}\label{sec:fixedscenario}
The first simulation study considered is one in which true response rates in scenarios are fixed with each basket having either a null ineffective response ($p_k=0.2$), effective response ($p_k=0.4$) or marginally effective response rate ($p_k=0.3$). The data scenarios considered are presented in Table \ref{tab:scenarios}, covering a wide range of cases. Scenario 1 is the global null state under which all baskets are ineffective to treatment, whereas, scenario 6 is the case where all baskets are truly effective to treatment. Scenarios 2-5 are cases in which the new basket is ineffective with a varying number of effective existing baskets. Similarly, scenarios 7-10 are cases in which the new basket is effective to treatment, again with a varying number of effective existing baskets. Scenarios 11-16 cover a variety of cases where baskets are a combination of ineffective, effective and marginally effective.

\begin{table}[ht]
\centering
\caption{Simulation study scenarios}\label{tab:scenarios}
\begin{tabular}[t]{lcccccclccccc}
\toprule
&$p_1$&$p_2$&$p_3$&$p_4$&$p_5$&&&$p_1$&$p_2$&$p_3$&$p_4$&$p_5$\\
\midrule
\text{Scenario 1} & 0.2 & 0.2 & 0.2 & 0.2 & 0.2 & & \text{Scenario 9} & 0.4 & 0.4 & 0.2 & 0.2 & 0.4\\
\text{Scenario 2} & 0.4 & 0.2 & 0.2 & 0.2 & 0.2 & & \text{Scenario 10} & 0.4 & 0.4 & 0.4 & 0.2 & 0.4\\
\text{Scenario 3} & 0.4 & 0.4 & 0.2 & 0.2 & 0.2 & & \text{Scenario 11} & 0.3 & 0.2 & 0.2 & 0.2 & 0.2\\
\text{Scenario 4} & 0.4 & 0.4 & 0.4 & 0.2 & 0.2 & & \text{Scenario 12} & 0.3 & 0.3 & 0.2 & 0.2 & 0.2\\
\text{Scenario 5} & 0.4 & 0.4 & 0.4 & 0.4 & 0.2 & & \text{Scenario 13} & 0.3 & 0.2 & 0.2 & 0.2 & 0.3\\
\text{Scenario 6} & 0.4 & 0.4 & 0.4 & 0.4 & 0.4 & & \text{Scenario 14} & 0.3 & 0.3 & 0.2 & 0.2 & 0.3\\
\text{Scenario 7} & 0.2 & 0.2 & 0.2 & 0.2 & 0.4 & & \text{Scenario 15} & 0.4 & 0.3 & 0.2 & 0.2 & 0.3\\
\text{Scenario 8} & 0.4 & 0.2 & 0.2 & 0.2 & 0.4 & & \text{Scenario 16} & 0.4 & 0.3 & 0.3 & 0.2 & 0.3\\
\hline
\end{tabular}
\end{table}%

The cut-off values $\Delta_{k_0}$ and $\Delta_{k^{\prime}}$ are calibrated for each approach separately as described in Table \ref{tab:approachesforaddingcondensed} in order to achieve a type I error rate of 10\% either under the null or across data scenarios using the RCaP. Note that under both calibration procedures, UNPL differs from the other 3 approaches as it's calibration only takes into account the $K_0=4$ existing baskets, with the new basket being an unplanned addition, thus calibration will occur given the following four scenarios $\bm{p_1}=(0.2,0.2,0.2,0.2),\;\bm{p_2}=(0.4,0.2,0.2,0.2),\;\bm{p_3}=(0.4,0.4,0.2,0.2)$ and $\bm{p_4}=(0.4,0.4,0.4,0.2)$ which cover all global and partial nulls given $K=4$ baskets of equal sample size.  

For IND, PL1 and PL2, under the traditional calibration procedure, calibration is conducted under scenario 1, the global null, to control error rates to the nominal 10\% level in this single scenario. However, the RCaP requires several scenarios to calibrate across. Consideration must be taken into which scenarios to include in calibration, here the assumption is made that all scenarios carry the same importance and thus weights were set as $\omega_m=1$ for all scenarios. Scenarios 1-5 cover all global and partial null cases considering just the existing baskets when the new basket is null and scenarios 7-10 again cover all global and partial null cases but when the new basket is non-null. (For this calibration cases of marginally effective response rates were not considered but may be incorporated based on clinicians beliefs). If sample sizes were equal across all $K$ baskets, scenarios 1-5 would sufficiently cover all global and partial nulls. Thus two options are considered: calibrate over just scenarios 1-5 or calibrate over all global and partial null scenarios 1-10 to take into account the unequal sample size in the new basket. A simulation study is presented in the supplementary material that compares these two options. Results indicated minimal differences in power and error rates and thus calibration across fewer scenarios is preferred due to the lower computational cost.

\begin{table}[ht]
    \centering
    \caption{Calibrated $\Delta_{k_0}$ and $\Delta_{k^{\prime}}$ values for IND, UNPL, PL1(a) and PL2(a) under the two separate calibration methods: calibration under the global null and the RCaP across scenarios 1-5.}\label{tab:cut_off}
    \begin{tabular}[t]{lcccccc}
    \toprule
         &  \multicolumn{4}{c}{Calibrate under the null} & \multicolumn{2}{c}{RCaP}\\
         & &$\Delta_{k_0}$ & $\Delta_{k^{\prime}}$ & &$\Delta_{k_0}$ & $\Delta_{k^{\prime}}$\\
         \midrule
         IND & &0.8599 & 0.8998 & &0.9030 & 0.8989\\
         UNPL & &0.8599 & 0.8599 & &0.9030 & 0.9030\\
         PL1(a) & &0.8566 & 0.8409 && 0.9034 & 0.9021\\
         PL2(a) & &0.8599 & 0.8409 & &0.9030 & 0.9021\\
         \hline
    \end{tabular}
\end{table}

Due to the unequal sample sizes across the new and existing baskets, for IND, PL1 and PL2, following Algorithm \ref{alg:cal}, $\Delta_{k_0}$ is selected as the 90\% quantile of the posterior probabilities in basket 4 across scenarios 1-4 in which it's true response is null, with $\Delta_{k^{\prime}}$ selected based on all scenarios 1-5. Cut-off values for calibration approaches are presented in Table \ref{tab:cut_off}.

\subsection{Fixed Scenario Simulation Results - A Comparison of Calibration Approach}\label{sec:SimCalibrationApproaches}
Taking this fixed data scenario simulation setting in which 16 fixed response rate data scenarios were considered, comparisons are first drawn between the traditional approach of calibrating cut-off values $\Delta_{k_0}$ and $\Delta_{k^{\prime}}$ to achieve a nominal error rate and the RCaP, where the cut-off values are calibrated across several scenarios to achieve an average nominal error rate over these scenarios. For the RCaP, calibration occurred across scenarios 1-5 for reasons previously stated. 

\begin{figure}[ht!]
    \centering 
    \includegraphics[width=1\textwidth]{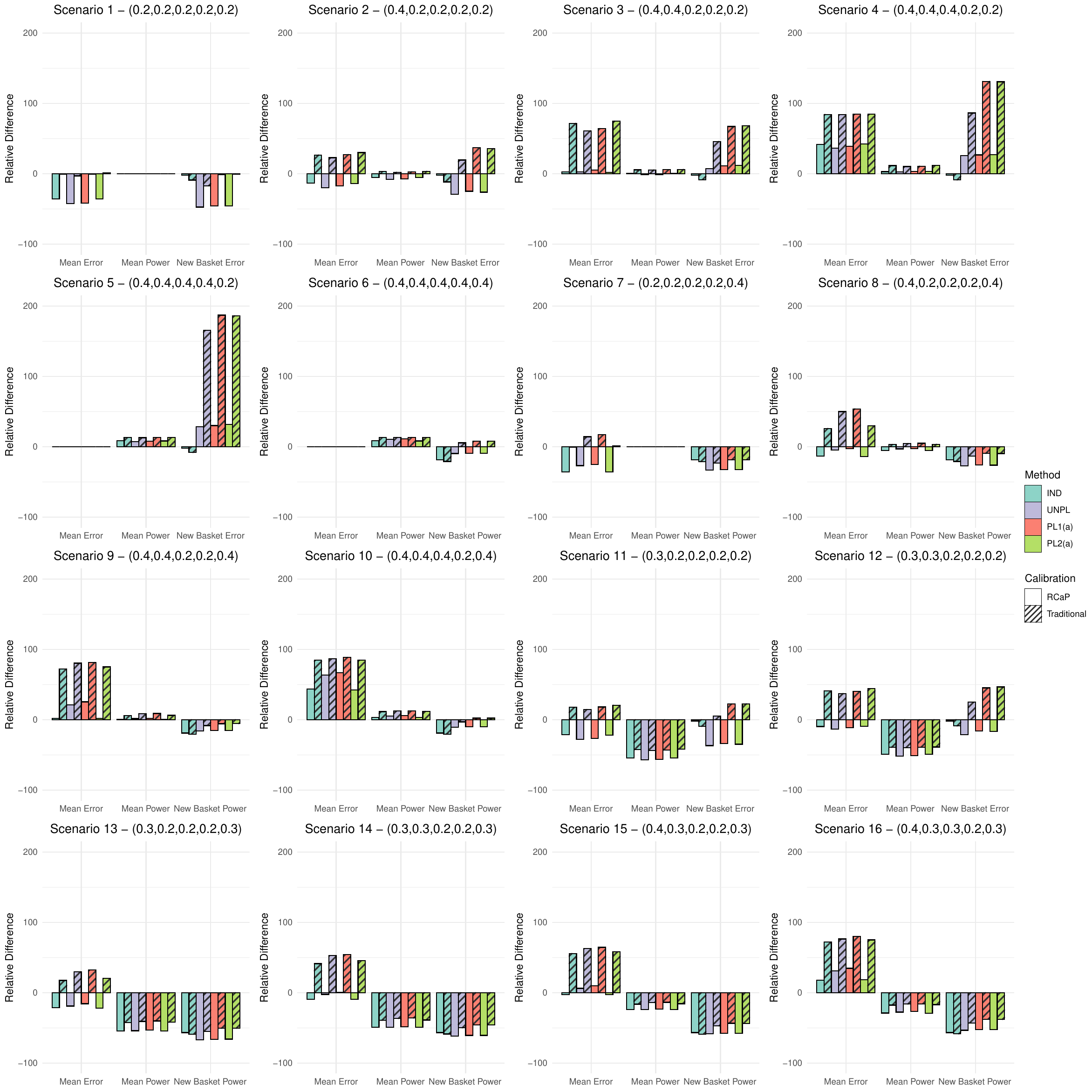}
    \caption{The relative difference in type I error rate and power compared to the targeted values of 10\% and 80\% respectively. This is given for all four approaches for adding a basket under the two different calibration schemes, the traditional calibration under the global null and the RCaP across scenarios 1-5. Results are split into 3 categories: mean error in which the percentage of data sets within which the null was rejected is averaged across all ineffective existing baskets; mean power as above but for all effective existing baskets and new basket error/power in which results are the percentage of data sets within which the null was rejected just in the new basket.}
    \label{fig:RelativeDiff}
\end{figure}

For each of the 16 fixed scenarios and 4 approaches for the addition of a basket, the relative difference between the observed type I error rate/power and targeted level (10\% and 80\% respectively) are measured under each calibration approach. These relative differences are presented in Figure \ref{fig:RelativeDiff}.

First consider cases in which the new basket has a null response rate, i.e. scenarios 1-5. Under the null scenario, the traditional calibration approach achieves exactly (with some simulation error) the nominal 10\% error rate, whilst the RCaP reduces the error rate up to 42.5\% of the nominal level in existing baskets and 47.4\% in the new basket. Under scenario 2, calibration across scenarios results in an under-powered study, with up to a 7.8\% reduction of the nominal 80\% level, however, this came with a 20.1\% decrease in type I error rate from the targeted value in existing baskets and 29.3\% in the new. Whereas, calibration under the null inflates the error rate by up to 30.1\% and 36.8\% in existing and new baskets respectively with a fairly similar power to the RCaP, although, slightly over the nominal level.  Similar findings hold in scenarios 3 and 4 in terms of error rates under a null calibration approach, however, error rates under the RCaP are also inflated but to a much lesser extent (e.g. 42.4\% compared to 84.7\% in existing baskets under PL2(a)) compared to the traditional calibration.  

The most blatant benefit of the RCaP is observed under scenario 5 in which the new basket is the only one with an ineffective response rate. For this basket, under a traditional calibration, error rates are almost tripled with values inflated by up to 186.1\% of the nominal 10\% level, compared to just 31.7\% under the RCaP. Under both calibrations, the study is slightly over-powered. Thus, calibrating across scenarios can massively reduce error rates in the new basket without issues in power arising in the existing baskets. 

In cases where the new basket is effective (scenarios 6-10), both calibration approaches tend to lead to under-powered estimates in the new basket with the exception of scenario 6 where the power in the new baskets is increased up to 8\% of the 80\% targeted value across the latter three approaches (IND, PL1 and PL2) under calibration under the null. In all scenarios, bar scenario 8, there are no issues in terms of power in the existing baskets, with all values greater than the nominal 80\% level, with slightly higher power observed under a traditional calibration approach. In scenario 8, power is up to a 5.4\% reduction of the nominal level using the RCaP compared to an increase of 3.2\% under a traditional approach. Error rates in existing baskets tends to increase as the number of effective existing baskets increases, with far higher error rates occurring under a traditional calibrate under the null approach. For example, under scenario 9, error rates are increased by up to 81.4\% of the nominal 10\% level compared to a 20.9\% increase under the RCaP. 

Scenarios 11, 12, 13 and 14 are parallel to the results under scenarios 2, 3, 9 and 10 respectively. The main differences lying in the results of power, as the baskets now have a marginally effective response rate, making it more difficult to distinguish between an effective and ineffective treatment effect. In all cases, power under the RCaP is lower due to the more conservative $\Delta_k$ cut-off values, however, this came with a reduced error rate across all 4 of these scenarios and 4 approaches. For instance, under scenario 14 error rates under the traditional calibration have a relative increase of 54\% over the nominal 10\% level, compared to error rates controlled at or below 10\% under the RCaP. Similar results are found under scenarios 14-16 in which baskets are a combination of effective, marginally effective and ineffective. 

From these findings, noticeable benefits can be identified of using a calibration across data scenarios compared to calibration under just the null. Across scenarios 1-10, in which all baskets are either effective or ineffective, estimates in existing baskets are only under-powered in two cases (scenarios 8 and 10) with a maximum reduction in power of 7.8\%. Whilst in the new basket, power tends to be below the nominal 80\% level under both calibration approaches due to the smaller sample size. More reduction is observed under the RCaP by up to 33.2\% compared to 23.2\% under the traditional approach. But this comes with massive reduction in error rates across all baskets. For existing baskets, under the traditional approach, error rates have up to a 88.6\% inflation over the nominal 10\% level compared to 66.7\% under the RCaP. Even more substantial reductions in error for the RCaP compared to the traditional approach are observed in the new basket with its reduced sample sizes. Error rates have up to an 186.1\% increase over the nominal 10\% level under the traditional approach with just a maximum increase of 31.7\% observed under the proposed calibration. 

Based on these results of clear benefits in type I error control, further results presented in this paper utilizes the RCaP in which, cut-off values $\Delta_{k_0}$ and $\Delta_{k^{\prime}}$ are calibrated across scenarios 1-5. However, results for simulation studies under the traditional calibration approach are provided in the supplementary material. 

\subsection{Fixed Scenario Simulation Results - A Comparison of Approaches for Adding a Basket}\label{sec:SimMethodComparison} 
\begin{figure}[ht!]
    \centering 
    \includegraphics[width=1\textwidth]{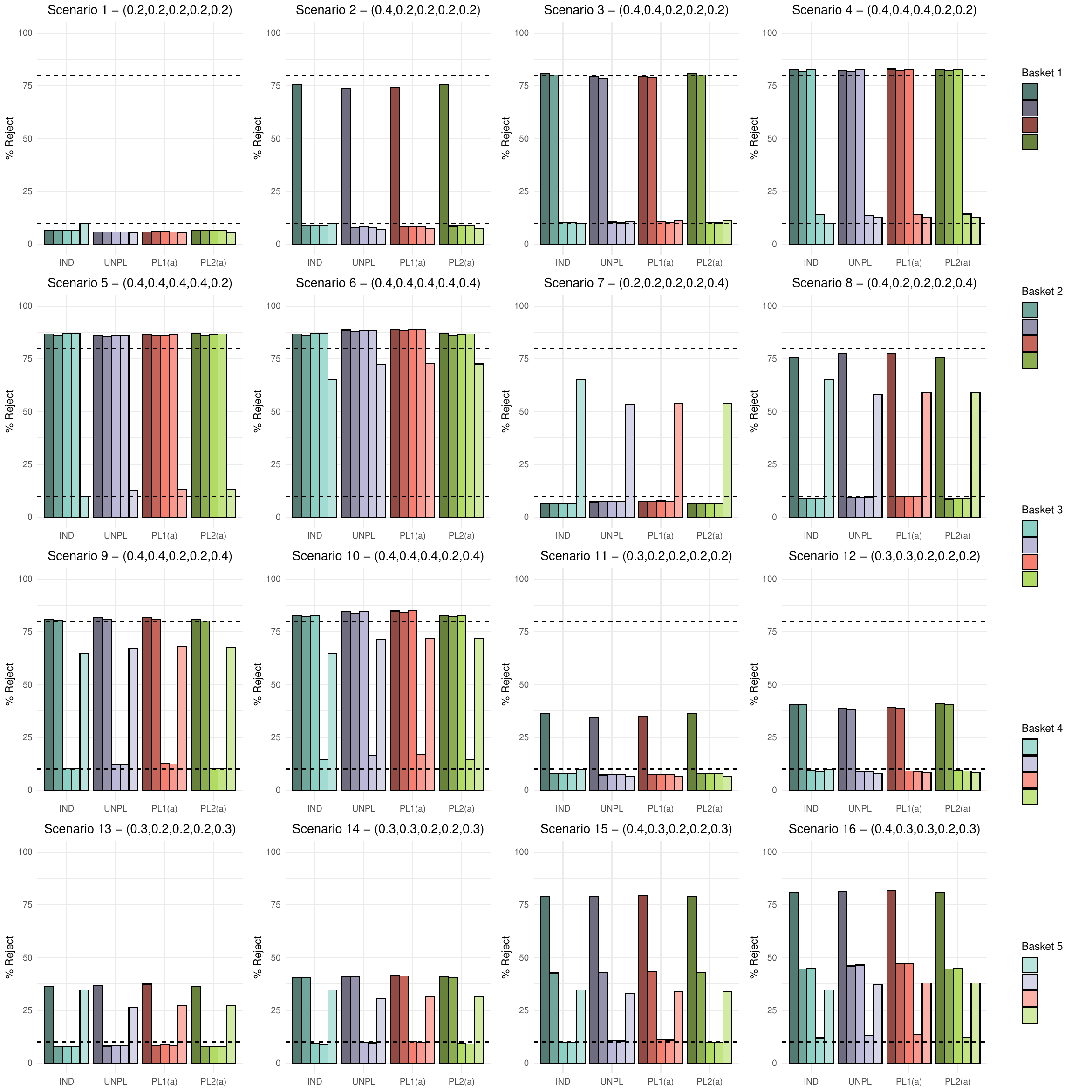}
    \caption{Fixed scenario simulation study results: The percentage of data sets within which the null hypothesis was rejected, where $\Delta_{k_0}$ and $\Delta_{k^{\prime}}$ were calibrated across scenarios 1-5 to achieve a 10\% type I error rate on average. This is plotted for each of the four approaches for adding a basket in all five baskets.}
    \label{fig:FixedScenariosAcross}
\end{figure}

Now consider the four approaches for the addition of a basket to an ongoing study under the fixed data scenario setting as previously described. The results for power and type I error rate for each approach under the 16 fixed data scenarios are presented in Figure \ref{fig:FixedScenariosAcross}, which show the percentage of simulated data sets in which the null hypothesis was rejected for each of the four approaches for adding a basket. Dashed lines represent both the nominal 10\% type I error rate and 80\% power.

As $\Delta_{k_0}$ and $\Delta_{k^{\prime}}$ are calibrated across scenarios 1-5 to achieve an average 10\% type I error rate, in some scenarios - including the global null case - the type I error rate lies below the nominal level. However, under IND, the new basket is always analyzed independently and as such, the error rate will remain the same across scenarios. This will lead to the new basket always having an error rate of approximately 10\%.

Under the global null, for existing baskets, the UNPL and PL1(a) approach in which information is borrowed between all $K=5$ baskets, including the homogeneous new basket, results in slightly lower type I error rates compared to other approaches at approximately 5.8\%. For the new basket, IND has substantially higher error rate which lies exactly at the nominal 10\% level. Approaches UNPL, PL1(a) and PL2(a) all have similar error rates in the new basket at around 5.3\%. 

When analysing existing baskets, IND and PL2(a) are equivalent as they both borrow via the EXNEX model between just the four existing baskets. This is observed in Scenarios 2 with both approaches giving the highest power at 75.7\%, which does lie below the targeted 80\% value, but is higher than UNPL and PL1 with power of 73.7\% and 74.1\% respectively. Both UNPL and PL1 borrow from the new basket when analysing the existing and hence, as the new basket has a null response rate, the posterior probabilities are pulled down towards the common mean, resulting in a higher number of type II errors. Error rates for all baskets are consistent across approaches with the exception of the IND approach where the new basket type I error is approximately 3\% higher. Similar observations are made in scenario 3 in terms of existing baskets, however, error in the new basket under IND is now the smallest, but the maximum inflation of error over the 10\% nominal level is only 1.1\% (under PL1(a) and PL2(a), which are equivalent when analysing new baskets).

Scenarios 4 and 5 show consistent power in all non-null existing baskets across all 4 approaches, all above the targeted 80\% level. The UNPL approach demonstrates marginally lower power than other methods. The average power under UNPL is 85.7\% compared to 86.2\% under PL1 in scenario 5. Both approaches analyze baskets in the same way, borrowing between all $K$ baskets via the EXNEX model, the only difference being the calibration approach. This led to slightly more conservative $\Delta_{k_0}$ value under UNPL compared to PL1(a), leading to fewer rejections of the null hypothesis and lower power/error rates. Under scenario 4, for basket 4, error rates are slightly higher under IND and PL2(a). This is due to the common mean across baskets 1-4 being higher than across all baskets, due to the new basket being ineffective to treatment. Hence, fewer false rejections should be made under UNPL and PL1(a). In terms of error rates in the new basket, PL1(a) and PL2(a) have marginally higher error rates at 13.1\% under scenario 5. As $\Delta_{k^{\prime}}$ is lower in PL1(a) compared to the UNPL approach, error is higher at 13\% compared to 12.8\%.

When the new basket is effective, under scenarios 6, 9 and 10, substantial improvements in power are observed in the new basket when information borrowing is utilized. Under scenario 6, PL1(a) gives the greatest power for all $K$ baskets. Without information borrowing, the reduced sample size in basket 5 results in substantially lower power at 65.0\% under an IND approach compared to 72.3\% under PL1(a). Similarly, in scenarios 9 and 10, as a number of existing baskets are also effective, borrowing information between all baskets substantially improves power in the new. Under the IND and PL1(a) approaches, error rates in the existing baskets are slightly higher at around 12.5\% in scenario 9 and 16.5\% in scenario 10. Whereas, error rates in IND and PL2(a) have an error rate of 14.3\% at the cost of reduced power in other baskets. 

However, when the new basket is effective and either none or one of the existing baskets is also effective (scenarios 7 and 8), due to the level of heterogeneity across baskets, borrowing information reduces power. In which case, analysis under the IND approach is more appropriate. 

Results of scenarios 11, 12 and 13 correlate to those of scenarios 2, 3 and 8 respectively but with marginally effective rather than effective true response rates. This gives much lower power across baskets but similar patterns in results. Scenario 14 does differ from the results of scenario 9, with IND now producing the highest power in the new basket.

Scenarios 15 and 16 have a combination of effective, marginally effective and ineffective baskets. For existing baskets, those that are marginally effective have similar power values across all approaches under scenario 15 of around 42.7\%, however, more variation is observed under scenario 16 with PL1(a) producing a power of 47\%, which is higher compared to UNPL with power 46.1\% and IND and PL2(a) at 44.7\%. But for the single effective basket all approaches give similar power values ranging from 78.7\%-79.0\% under scenario 15 and 80.9\%-81.8\% under scenario 16. Error rates in the existing baskets are higher under the IND and PL1(a) approaches for both scenarios as the posterior probabilities are pulled up via borrowing from the effective new baskets. The maximum error rate across both scenarios is 13.5\%. For the new basket UNPL, PL1 and PL2 are almost identical in power, with the IND approach giving lower power under scenario 16 due to the lack of borrowing from the mostly homogeneous existing baskets. 

Combining these findings, when the new basket is null, power in existing baskets tends to be slightly lower in the UNPL and PL1 approach which borrow from this ineffective basket. Saying this, the largest difference in power across approaches is just 2\% (which occurs in scenario 2). The IND and PL2(a) approaches tend to demonstrate a fraction more error inflation for existing baskets, with IND also the least favourable in the new basket with error rates always lying at the nominal 10\% level due to the independent analysis. All 3 UNPL, PL1 and PL2 approaches have similar error rates in the new basket. When the new basket is effective, performance of all approaches varies based on the number of effective existing baskets. When this number is high substantial power can be gained in the new basket from borrowing between all baskets in the trial. However, when very few existing baskets are effective, borrowing actually reduces power in this basket compared to analysing independently. Across all cases, in existing baskets, UNPL and PL1(a) demonstrate both higher power and error compared to IND and PL2(a) due to borrowing from the effective new basket. 

\subsection{Random Scenarios Simulation}\label{sec:SimVariedTruth}
Based on the fixed response rate simulation results presented in the previous study, no one approach is clearly the most appropriate for use, with little difference observed between some approaches. Therefore a further simulation study was conducted in which, rather than fixing the true response rate for the new basket prior to the trial, it is randomly generated within each trial run of the simulation. The goal of such a simulation study is to distinguish were discrepancies between approaches arise.

Following the same set-up as the fixed scenario case with four existing and one new basket, four settings were considered. In each setting the response rates for existing baskets are fixed while the response rate for the new basket is randomly selected with uniform probability across an interval. Three sub-cases are considered in each setting, varying the interval from which $p_5$ is sampled: sub-case (a) in which the new basket is ineffective to treatment so $p_5\in[0.1,0.2]$, sub-case (b) in which the new basket has an effective response rate so $p_5\in[0.4,0.5]$ and finally sub-case (c) in which the new basket is marginally effective to treatment so $p_5\in[0.2,0.3]$. The four settings are:

\begin{enumerate}
    \item Fix the response rate in all the existing baskets as ineffective, i.e. $p_{1,2,3,4}=0.2$, with $p_5$ varied across one of the 3 intervals (a), (b) or (c). \item Fix the response rate in all the existing baskets as effective, i.e. $p_{1,2,3,4}=0.4$, with $p_5$ varied across one of the 3 intervals (a), (b) or (c).
    \item Fix the response rate in two of the existing baskets as effective, i.e $p_{1,2}=0.4$ and two ineffective, i.e. $p_{3,4}=0.2$, with $p_5$ varied across one of the three intervals (a), (b) or (c).
    \item Fix the response rate in one of the existing baskets as effective ($p_1=0.4$), two as marginally effective ($p_{2,3}=0.3$) and one as ineffective ($p_5=0.2$), with $p_5$ varied across one of the 3 intervals (a), (b) or (c).
\end{enumerate}

Calibrated $\Delta_{k_0}$ and $\Delta_{k^{\prime}}$ values used for hypothesis rejection are the same as that in Table \ref{tab:cut_off} where calibration is conducted using the RCaP across scenarios  1-5.

A total of 12 simulation settings were considered (the four settings outlined above under each of the 3 sub-cases for sampling $p_5$) with 10,000 randomly generated data scenarios within each. In each sub-case of the four settings, pair-wise discrepancies between approaches were identified in terms of differing decisions regarding the rejection of the null hypothesis in a basket (and hence differing efficacy conclusions). 

\subsection{Random Truth Simulation Results}
\begin{figure}[ht]
    \centering \hspace{-2em}
    \includegraphics[width=1\textwidth]{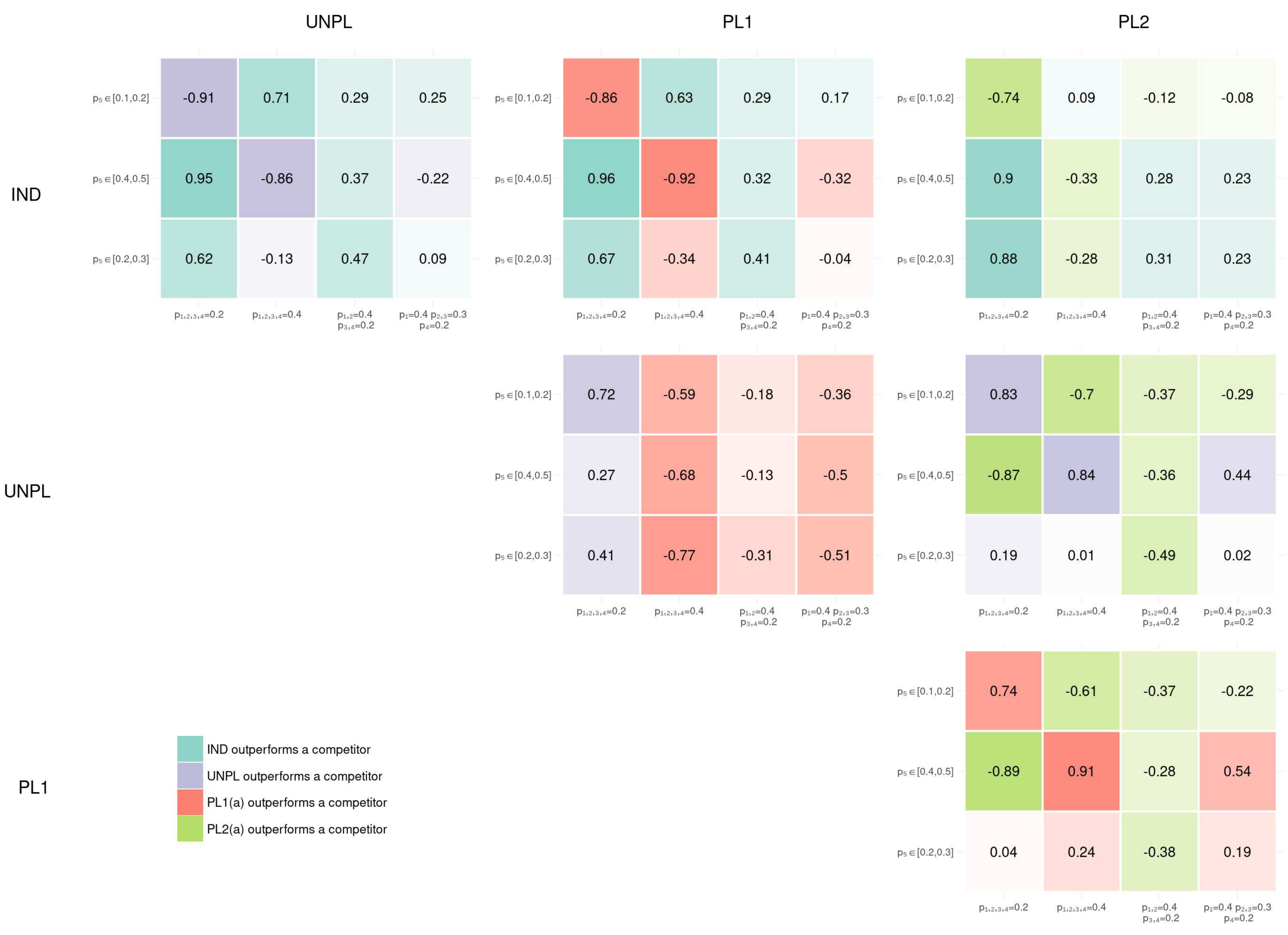}
    \caption{Pair-wise comparison between approaches in each of the 12 simulation settings within which the true response rate in the new basket is varied. The heat map presents the difference in proportion of times the approach corresponding to rows outperformed the approach corresponding to the column (with negative values indicating the approach in the column gave more correct conclusions over the approach in the row where discrepancies between the two approaches arise). The color in the heat map represents which approach gave superior correct conclusion, with shade representing the amount of difference between approaches. Blue represents IND giving more correct conclusions where discrepancies lie, Purple for UNPL, Red for PL1 and Green for PL2.}
    \label{fig:HeatMapVariedTruths}
\end{figure}

Results of the 72 pair-wise comparisons across the 12 simulation settings are plotted in Figure \ref{fig:HeatMapVariedTruths}. This looks at the difference in proportion of correct conclusions made where discrepancies between the two approaches arose. As an example of a discrepancy, consider setting 1 sub-case (a) in which existing baskets are null, $p_{1,2,3,4}=0.2$ and the new basket is also null with response rate randomly generated from the interval $p_5\in[0.1,0.2]$. A pair-wise discrepancy arises when two approaches give a different conclusion regarding whether or not a basket is effective to treatment, for instance when the IND approach states the treatment is effective in basket 5 but UNPL states it is ineffective. In this case as the treatment is in-fact ineffective, UNPL led to the correct conclusion thus outperforming IND in this simulation run. Discrepancies are detected across every basket in each of the 10,000 simulation settings and approach which gave the correct inference recorded. Proportions are then taken of correct conclusions in these discrepancies for both approaches under comparison. Going back to Figure \ref{fig:HeatMapVariedTruths}, each sub-plot represents a comparison between two approaches for each of the 12 simulation settings. A negative proportion implies the approach corresponding to the column outperformed the competitor approach in the corresponding row in terms of correct conclusions drawn where discrepancies occur. The colour in the sub-plot represents the superior approach with shade depicting the degree of difference in proportion between the two. 

First, consider the pair-wise comparison between IND and UNPL. The IND approach outperforms UNPL by making more correct conclusions in cases of discrepancies in 8 of the 12 simulations. In setting 1 where the existing baskets are null, the difference in approaches is substantial. When the new basket is effective in setting 1, IND is preferred giving the correct conclusion in 80.9-97\% of cases, but when ineffective, UNPL gives correct conclusions in 95.4\% of cases where discrepancies lie. Other cases where UNPL is preferred over IND is when there is again homogeneity between existing and new baskets, i.e. in setting 2 where both new and existing baskets are effective. When there is heterogeneity between all baskets, IND in which the new basket is analyzed independently tends to outperform the approach that utilizes an unplanned addition of a basket.

The analysis approach in UNPL is identical to that in PL1(a), the only difference being the calibrated $\Delta_{k_0}$ and $\Delta_{k^{\prime}}$ values obtained based on an unplanned and planned addition respectively. As such, a similar pattern in results to the IND/UNPL pair-wise comparison are obtained in the comparison between IND and PL1(a). Under UNPL, these cut-off values are more conservative, leading to fewer rejections of the null compared to PL1(a) regardless of whether a basket is truly effective or not. In all cases except setting 1, PL1(a) outperforms UNPL in terms of correct conclusions made. Under setting 1, when the existing baskets are all null, the ideal is for the hypothesis not to be rejected and thus the more conservative $\Delta_{k_0}$ value leads to more correct conclusions being made. Breaking down these results further and looking specifically at the new basket only, UNPL leads to correct conclusions in only 3.6-5.4\% of discrepancies under setting 1 when the new basket is effective. When ineffective, in all simulation runs, UNPL led to the correct conclusion when discrepancies were identified between the two approaches. So the superior performance in the existing baskets (81.3-85\% of correct conclusions) for UNPL in setting 1 overrides the poor performance in the new basket when it is effective. However, in cases where at least one existing basket is effective, PL1(a) gives better correct conclusions over UNPL. This will come from the less conservative cut-off values, leading to more correct rejections and hence higher power. From this we can determine that in most cases, a planned addition of a basket gives superior decisions over an unplanned addition.

Under the IND and PL2(a) approaches, analysis for existing baskets is equal and leads to the same conclusions, so the only discrepancies between rejections of the null will occur in the new basket. In settings 2-4 when at least one existing basket is effective, approaches are fairly equal in terms of difference in correct conclusions, with IND performing best when there is heterogeneity between all basket with the new basket effective (61.3-65.6\% simulation discrepancies where IND gave the correct conclusion over PL2(a)). The Pl2(a) approach has superior performance when all baskets are homogeneous as information borrowing between all baskets improves power and precision.  

Similarly, under PL1(a) and PL2(a) analysis for the new basket is equal and thus differences only lie in existing baskets. In cases of complete homogeneity between existing baskets with homogeneity also between the new basket, PL1(a) is the clear winner as precision can be gained through borrowing between all baskets. However, in cases where more heterogeneity is observed such as when the new basket is effective and existing ineffective and vice-versa, PL2(a) is superior as it does not draw on information from these heterogeneous baskets when analysing existing baskets. The comparisons between UNPL and PL2(a) result in the same conclusions. 

In summary, as in the fixed response rate case, these results do not highlight a clear `winner' in terms of optimal approach implemented in the addition of a new basket. However, key findings can be drawn from these results such as the finding that in most cases, a planned addition of a new basket outperforms an unplanned addition when information is borrowed between all baskets. We also note that analyzing the new basket as independent outperforms its competitor approach in 22 out of 36 comparisons, but in particular when heterogeneity is observed. The PL1(a) outperforms IND in the same cases in which PL1(a) also outperforms PL2(a) (i.e. homogeneity between new and existing baskets), else analyzing the new basket as independent is appropriate, in which case one would opt for the IND approach. In cases where PL2(a) outperforms PL1(a), IND outperformed or performed similarly to PL2(a). From this we then determine that PL1(a) tends to be a better approach than PL2(a) (and IND) in cases of homogeneity, with IND the optimal choice otherwise. 

\section{Simulation Study - Investigating the Robustness to the Timing of Addition}\label{sec:SimTiming}
In previous simulation studies it is assumed that the timing of addition of a new basket is known prior to the trial, however, this could easily not be the case. One would like to determine the effect of timing of addition, and thus sample size in new baskets, on the performance of analysis methods. 

\begin{figure}[ht]
    \centering 
    \includegraphics[width=1\textwidth]{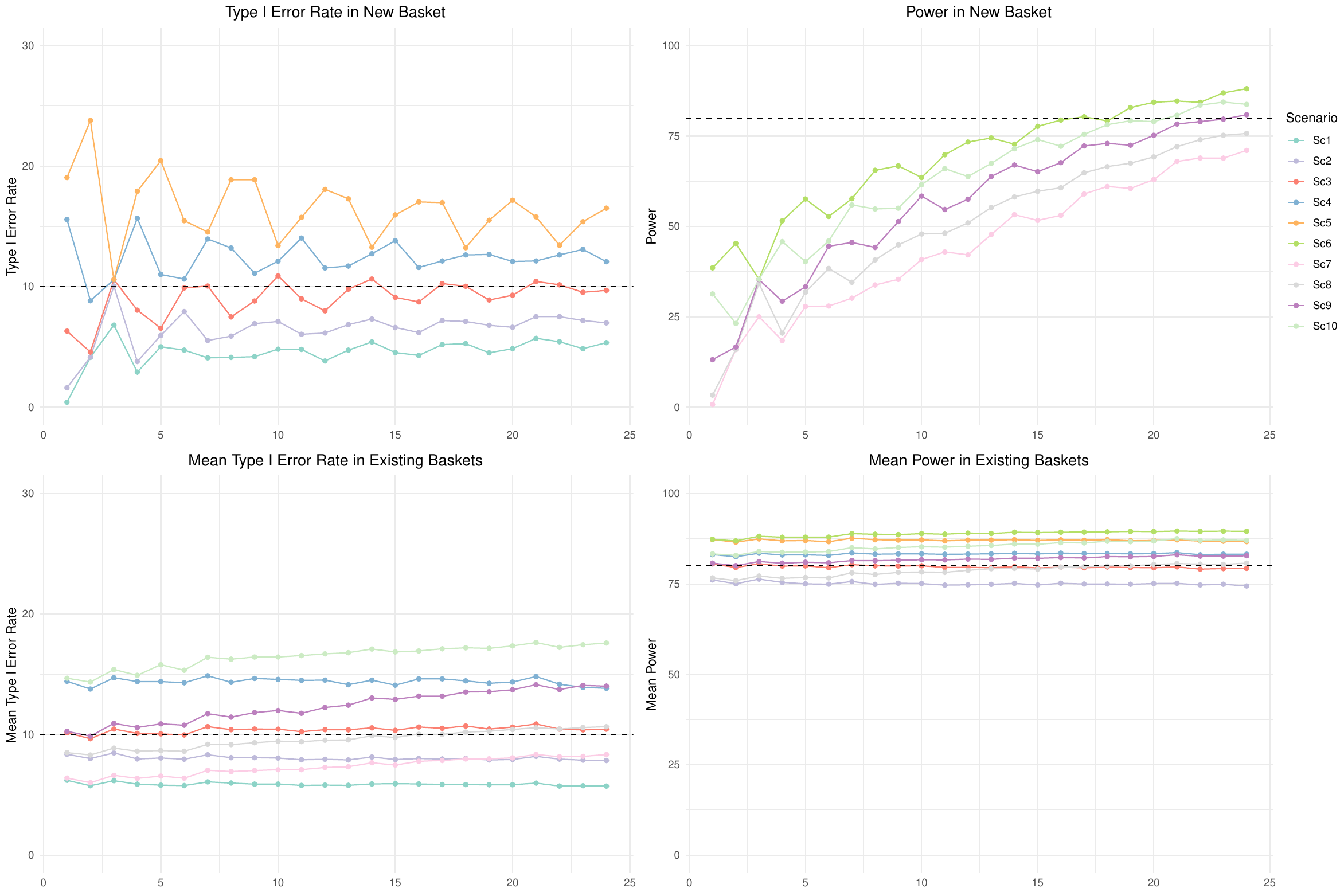}
    \caption{Type I error rate and power under each sample size of $n_5$ from 1 to 24 by applying PL1, split by existing and new baskets.}
    \label{fig:worstcase3b}
\end{figure}

When a basket is added later in the trial, sample sizes are smaller and thus the benefits of information borrowing increase. This means that methods which utilize such approaches become increasingly more superior over the IND approach in which an independent analysis is conducted. However, improvements in performance can still be obtained via information borrowing when addition of the basket is conducted early on in the trial.

Under IND, timing of addition will not have an impact on existing baskets due to the independent analysis of new baskets. Thus, when sample sizes are smaller the only effect will be reduced power in the new basket with increased power as sample sizes grow. Under UNPL, the addition is not planned and so timing of addition has no relevance to the calibration procedure. See the supplementary material for further exploration of the effect of sample size on the performance of IND and UNPL. 

Both PL1 and PL2 make a planned addition of a basket whilst utilizing information borrowing so timing of addition is taken into account in both the calibration process and analysis. In the likelihood that the sample size of the new baskets is unknown, performance of these approaches is more liable to change and thus the robustness of the approaches to the timing of addition is now explored. 

\begin{figure}[ht]
    \centering 
    \includegraphics[width=1\textwidth]{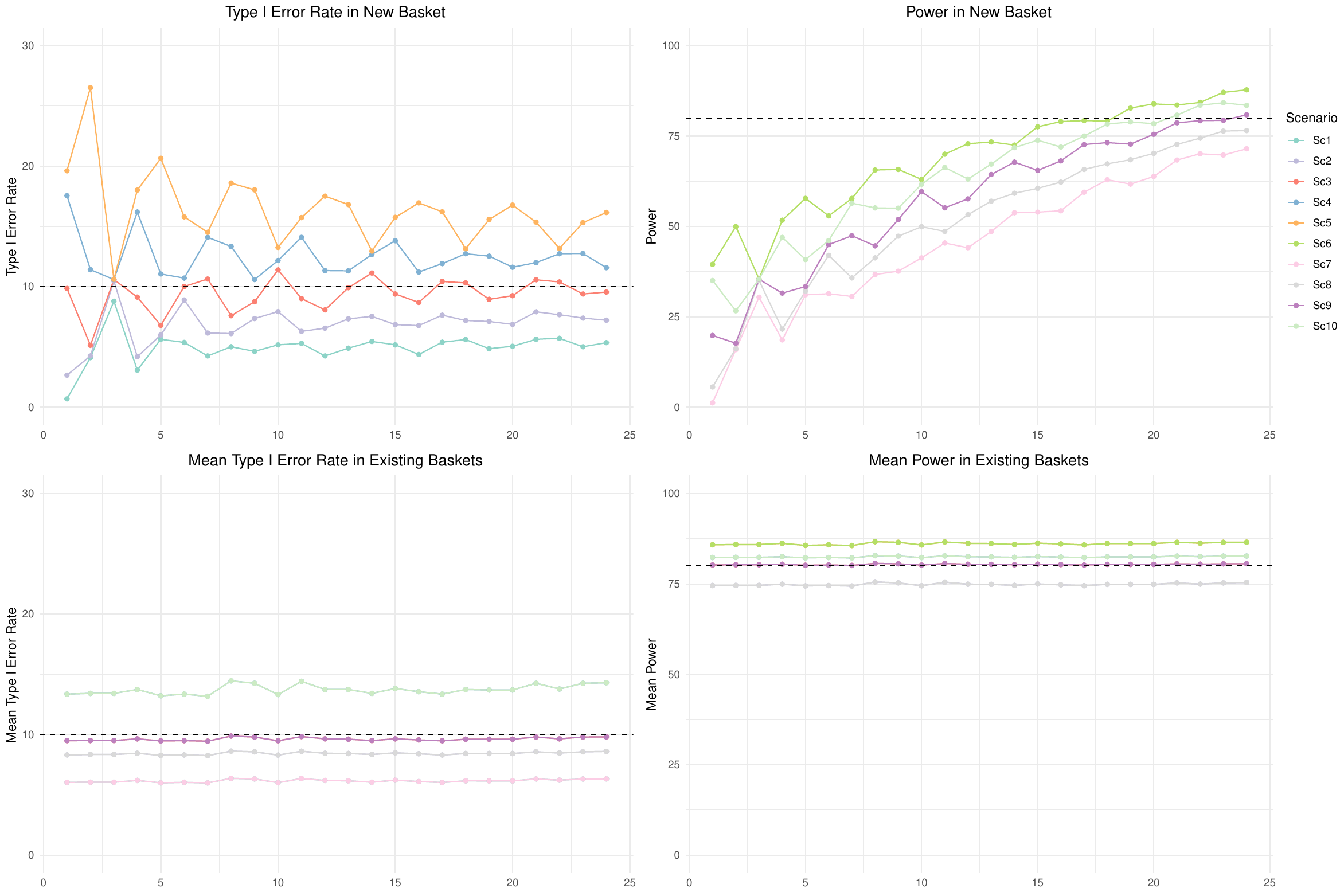}
    \caption{Type I error rate and power under each sample size of $n_5$ from 1 to 24 by applying PL2, split by existing and new baskets.}
    \label{fig:worstcase4b}
\end{figure}

To do so, again consider the fixed data scenario simulations setting with four existing and 1 new basket calibrated using the RCaP. In the previous simulation study, sample size of the new basket is assumed as known, consisting of $n_5=14$ patients, whilst existing baskets had $n_{k_0}=24$ patients in each. Now assume $n_5$ is unknown.

First consider PL1(b) applied to all possible sample sizes from $n_5=1$ up to the full sample size of existing baskets, $n_5=24$, with separate calibrations of $\Delta_{k_0}$ and $\Delta_{k^{\prime}}$ conducted for each value of $n_5$. Figure \ref{fig:worstcase3b} presents the type I error rate and power in new and existing baskets for each value of $n_5=1,\ldots,24$ under scenarios 1-10 as presented in Table \ref{tab:scenarios}.

Error rates and power for existing baskets are fairly consistent across all sample sizes, implying the time of addition of a new basket has little to no impact on the performance in these existing baskets, obviously an ideal characteristic. 

However, much more noticeable changes are observed in the new basket. Cyclic fluctuations occur in both power and error rate due to the discreetness of data. As expected, power in the new basket generally increases with the sample size as more information is available, hence improving precision of posterior distributions. Across scenarios, power increases as the number of effective existing baskets also increases. The targeted 80\% level is only reached under scenario 6 when $n_5=18$ and $n_5=21$ under scenario 7. The nominal level is never achieved under scenarios 7 and 8 in which there are none or just one effective existing basket.

In terms of error rates, more variation tends to occur when sample sizes are small, with the greatest error occurring when there are just 2 patients in the new basket (error of 23.8\%) under scenario 5. As the number of effective existing baskets increases, the error rates are uniformly higher, with scenario 5 as the `worst case' scenario where the only ineffective basket is the new basket. However, there is no general increase or decrease as the sample size increases and thus one cannot make the conclusion that any one sample size results in a more detrimental performance, at least in terms of the type I error rate. 

The same interpretations are drawn when looking at the timing of addition under PL2 as plotted in Figure \ref{fig:worstcase4b}. As PL1 and PL2 are equivalent for the new basket, with information borrowed between all baskets, the plots for type I error rate and power in the new basket are identical to that in Figure \ref{fig:worstcase3b}. Again, little to no variation is present in error and power for existing baskets as $n_5$ changes.

In summary, operating characteristics are fairly robust to the timing of addition of the new basket, particularly in the case of existing baskets, with little to no changes in power and error rate with the variation of sample sizes in the new basket. Power in the new basket is obviously improved as the sample size increases but no increase/decrease outside of the cyclic behaviour is observed in error rates, implying the type I error rate will be fairly unaffected by the timing of addition. 

\section{Discussion}
In this paper, we present four approaches for calibration and analysis of trials when a new basket is added part-way through. Approaches utilize the EXNEX Bayesian information borrowing model which was selected for its flexible borrowing between subsets of baskets. Other information borrowing models in the literature, such as the BHM \citep{Hierarchical}, CBMH \citep{cbhm}, MUCE design \citep{MUCE}, Liu's two-path approach \citep{Liu2}, RoBoT design \citep{RoBot} and a Bayesian model averaging approach \citep{BMA} are not considered here. However, we would expect the approaches in which information is borrowed between all baskets to possess even more inflation in error rates in the presence of any heterogeneous baskets. 

Through the thorough simulation studies, no one of the outlined approaches for adding outperforms its competitors across all cases. An approach which analyses new baskets as independent whilst retaining information borrowing between existing baskets understandably has better error control and power in cases of heterogeneity between new and existing baskets. However, significant power can be gained via information borrowing between all baskets when the new basket is homogeneous to existing ones. Under such an approach it is also shown that results are fairly robust to the timing of addition, with increased power in new baskets when sample sizes are larger but consistent error and power in existing baskets. This implies the size of the new basket has no detrimental effect on baskets that opened at the commencement of the trial. As sample size increases, the difference in error rates/power between analysing the new basket as independent and conducting information borrowing, will decrease and thus in such a case it may be beneficial to always analyse as independent to avoid issues when heterogeneity arises. 

Although all simulation studies conducted consisted of a single basket being added alongside four existing baskets, a further simulation study with 2 existing and 2 new baskets is presented in the supplementary material. Results imply the same conclusions as drawn in the simulation studies presented in this paper, but with an unplanned addition performing significantly worse than other approaches due to the lack of certainty in the calibration process with only two relatively small baskets being used. It is believed that as the ratio of existing to new baskets increases, the power gained through information borrowing in the new basket further improves due to the gain in certainty around point estimates. 

We have also promoted a transition away from the traditional calibration approach in which the type I error rate is controlled under a global null scenario, towards the novel calibration technique, RCaP, presented in this paper in which the type I error rate is controlled across several plausible data scenarios. The calibration method provides flexibility by allowing the clinician to specify potential outcomes of the trial in which one would like to control the error rate across. Compared to the traditional approach, the RCaP demonstrates superior error control with only a small loss in power relative to the targeted value in a handful of cases. 

Other adaptive design features such as interim analyses with futility/efficacy stopping are desirable in such clinical trials and have been considered across the literature around information borrowing in basket trials including in the work by Jin et al\cite{Assumption}, Berry et al\cite{Hierarchical}, Chu and Yuan\cite{cbhm} and Psioda et al\cite{BMA}. No such design features were considered in this paper which may be a limitation. The methodology described here could be extended to incorporate such features and future work into this aspect is being conducted. 

\section*{Data Availability}
All simulations were conducted through the computing software JAGS in R through the `rjags' package \citep{jags}. No new data have been used in this publication. Simulations can be reproduced using the open accessible code available at https://github.com/LibbyDaniells/AddingABasket.

\section*{Acknowledgments}
    This report is independent research supported by the National Institute for Health Research (NIHR Advanced Fellowship, Dr Pavel Mozgunov, NIHR300576). The views expressed in this publication are those of the authors and not necessarily those of the NHS, the National Institute for Health Research or the Department of Health and Social Care (DHSC). T Jaki and P Mozgunov received funding from UK Medical Research Council (MC$\_$UU$\_$00002/14 and MC$\_$UU$\_$00002/19, respectively). This paper is based on work completed while L. Daniells was part of the EPSRC funded STOR-i centre for doctoral training (EP/S022252/1). For the purpose of open access, the author has applied a Creative Commons Attribution (CC BY) licence to any Author Accepted Manuscript version arising.

\section*{Supporting information}
The following supporting information is available as part of the online article: Full model specifications and simulation results for the fixed scenario case with additional metrics including: family-wise error rates, mean point estimates along with their standard deviations and the percentage of cases where the correct conclusion was made across all baskets. A comparison study is presented between simulation studies in which the cut-off values are calibrated across a different number of scenarios using the RCaP. Full results for a further simulation study are given where cut-off values were calibrated under the traditional approach under the global null. Additional results for the semi-random truth simulation are provided, with heat maps split by new and existing baskets. Exploration is made in terms of the robustness of timing of addition of a new basket under both IND and UNPL. Finally, a further simulation study is presented consisting of two existing plus two new baskets added.

\bibliographystyle{apalike}
\bibliography{References}

\newpage
\section*{Supplementary Materials}

\setcounter{section}{0}
\setcounter{equation}{0}
\setcounter{figure}{0}
\setcounter{table}{0}
\setcounter{page}{1}
\makeatletter
\renewcommand{\theequation}{S\arabic{equation}}
\renewcommand{\thefigure}{S\arabic{figure}}
\renewcommand{\thesection}{SM\arabic{section}}
\renewcommand{\bibnumfmt}[1]{[S#1]}
\renewcommand{\citenumfont}[1]{S#1}
\renewcommand{\thetable}{S\arabic{table}}

\begin{table}[H]
\centering
\caption{Summary of approaches for analysis and calibration when adding a basket.}\label{tab:approachesforadding}\scalebox{0.85}{
\begin{tabular}{|P{1.5cm}|P{4.8cm}|P{4.8cm}|P{4.6cm}|}
\hline
Approach & Description & Calibration & Analysis\\
\hline
IND & Treat the new and existing baskets separately and independent of one another. Analysing the new basket as independent of existing baskets eliminates the potential negative effects of reduced information in the new baskets on existing baskets. & Calibrate $\Delta_{k_0}$ based on an EXNEX model applied to the $K_0$ existing baskets. For new baskets, calibrate $\Delta_{k\prime}$ based on either: (a) independent analysis conducted for each of the $K\prime$ new baskets or (b) borrow information between all $K\prime$ new baskets through a separate EXNEX model. & Analyse in the same way as calibration, with an EXNEX fitted to existing baskets and new baskets analyzed with an independent model.\\
\hline
UNPL & Naive approach in which an unplanned addition of new baskets is made and not considered in the calibration procedure. This occurs when it is unknown a basket will be added but it is then believed that borrowing information between all baskets, new and old, can improve inference. & Calibrate $\Delta_{k_0}$ based on an EXNEX model applied to the $K_0$ existing baskets. Fix $\Delta_{k\prime}=\Delta_{k_0}$ once new baskets are added. & Analyse by borrowing information between all $K$ baskets through an EXNEX model. \\
\hline
PL1 & It is known that a basket will be added during the study and information will be borrowed between all baskets new and existing. This may occur when it is apparent that a basket of patients will benefit from the study, however is not ready in time for the commencement of the trial and thus is planned to be added at a later time. & Calibrate $\Delta_{k_0}$ and $\Delta_{k\prime}$ based on an EXNEX model applied to all $K$ baskets. When (a) timing of addition is known: the sample sizes $n_k$ for all $K$ baskets are known and fixed in the EXNEX model. When (b) timing of addition is unknown: further simulation studies are required to explore the effect of $n_{k\prime}$ on operating characteristics, one could calibrate based on the least favourable configuration. & Analyse in the same way as calibration, with an EXNEX model fitted to all $K$ baskets. \\
\hline
PL2 & It is known that a basket will be added during the study but when conducting inference on existing baskets only information from other existing baskets is utilized, whereas for inference on new baskets, information is borrowed between all baskets in the trial. This will eliminate the effect of reduced sample sizes in new baskets on estimation of response rates in existing baskets whilst improving power and precision in the new basket. & Calibrate $\Delta_{k_0}$ based on an EXNEX model applied to just the $K_0$ existing baskets. Calibrate $\Delta_{k\prime}$ based on an EXNEX model applied to all $K$ baskets. When (a) the timing of addition is known, sample sizes, $n_k$, for all baskets are fixed in the calibration procedure. When (b) the timing of addition is unknown, further simulation studies would be required to explore the effect of $n_{k\prime}$ on operating characteristics and adjust calibration accordingly. & Analyse in the same way as calibration with an EXNEX model fitted to just the $K_0$ existing baskets when analysing existing baskets and with an EXNEX model fitted to all $K$ baskets when analysing the new basket(s). \\
\hline
\end{tabular}}
\end{table}

\section{Model Specification}\label{sec:ModelSpec}
The trial consists of a total of $K$ baskets, divided into $K_0$ existing baskets and $K\prime$ new baskets. Parameter choices are those implemented throughout the simulation studies presented in the main text.\\\\
\textbf{IND} Calibrate and analyse based on the following model:
\begin{center}\vspace{-1.5em}
\begin{minipage}[t]{.45\textwidth}
  \begin{align*}
         &Y_{k}\sim\text{Binomial}(n_k,p_k), &k&=1,\ldots,K\\
    &\theta_k=\log\Big(\frac{p_k}{1-p_k}\Big),\\
    &\theta_{k\prime}\;\sim\;\text{N}(-1.39,10^2), &k\prime&=K_0+1,\ldots,K\\
    &\theta_{k_0}=\delta_{k_0} M_{1k_0}+(1-\delta_{k_0})M_{2k_0},&k_0&=1,\ldots,K_0\\
  \end{align*}
\end{minipage}%
\begin{minipage}[t]{0.45\textwidth}
\begin{align*}
&\delta_{k_0}\sim\text{Bernoulli}(\pi_{k_0}),\\
    &M_{1k_0}\sim\text{N}(\mu,\sigma^2),\\
      &\mu\sim\text{N}(-1.39,10^2),\\
    &\sigma\sim \text{Half-Normal}(0,1),\\
     &M_{2k_0}\sim \text{N}(-0.85,4.76),
\end{align*}
    \end{minipage}
    \end{center}
with $\pi_{k_0}=0.5$ for all $k_0=1,\ldots,K_0$.\\\\

\noindent\textbf{UNPL} Calibrate based on the following model:
\begin{center}\vspace{-1.5em}
\begin{minipage}[t]{.45\textwidth}
  \begin{align*}
         &Y_{k_0}\sim\text{Binomial}(n_{k_0},p_{k_0}),&k_0&=1,\ldots,K_0\\
    &\theta_{k_0}=\log\Big(\frac{p_{k_0}}{1-p_{k_0}}\Big),\\
    &\theta_{k_0}=\delta_{k_0} M_{1k_0}+(1-\delta_{k_0})M_{2k_0},\\
    &\delta_{k_0}\sim\text{Bernoulli}(\pi_{k_0}),
  \end{align*}
\end{minipage}%
\begin{minipage}[t]{0.45\textwidth}
\begin{align*}
    &M_{1k_0}\sim\text{N}(\mu,\sigma^2),\\
      &\mu\sim\text{N}(-1.39,10^2),\\
    &\sigma\sim \text{Half-Normal}(0,1),\\
     &M_{2k_0}\sim \text{N}(-0.85,4.76),
\end{align*}
    \end{minipage}
    \end{center}
with $\pi_{k_0}=0.5$ for all $k_0=1,\ldots,K_0$. Analyse based on the following model:\\
\begin{center}\vspace{-2.5em}
\begin{minipage}[t]{.45\textwidth}
  \begin{align*}
         &Y_{k}\sim\text{Binomial}(n_k,p_k),&k&=1,\ldots,K\\
    &\theta_k=\log\Big(\frac{p_k}{1-p_k}\Big),\\
    &\theta_k=\delta_kM_{1k}+(1-\delta_k)M_{2k},\\
    &\delta_k\sim\text{Bernoulli}(\pi_k),
  \end{align*}
\end{minipage}%
\begin{minipage}[t]{0.45\textwidth}
\begin{align*}
    &M_{1k}\sim\text{N}(\mu,\sigma^2),\\
      &\mu\sim\text{N}(-1.39,10^2),\\
    &\sigma\sim \text{Half-Normal}(0,1),\\
     &M_{2k}\sim \text{N}(-0.85,4.76),
\end{align*}
    \end{minipage}
    \end{center}
with $\pi_k=0.5$ for all $k=1,\ldots,K$.\\\\

\noindent\textbf{PL1} Calibrate and analyse based on the following model:
\begin{center}\vspace{-1.5em}
\begin{minipage}[t]{.45\textwidth}
  \begin{align*}
         &Y_{k}\sim\text{Binomial}(n_k,p_k),&k&=1,\ldots,K\\
    &\theta_k=\log\Big(\frac{p_k}{1-p_k}\Big),\\
    &\theta_k=\delta_kM_{1k}+(1-\delta_k)M_{2k},\\
    &\delta_k\sim\text{Bernoulli}(\pi_k),
  \end{align*}
\end{minipage}%
\begin{minipage}[t]{0.45\textwidth}
\begin{align*}
    &M_{1k}\sim\text{N}(\mu,\sigma^2),\\
      &\mu\sim\text{N}(-1.39,10^2),\\
    &\sigma\sim \text{Half-Normal}(0,1),\\
     &M_{2k}\sim \text{N}(-0.85,4.76),
\end{align*}
    \end{minipage}
    \end{center}
with $\pi_k=0.5$ for all $k=1,\ldots,K$.\\\\

\noindent\textbf{PL2} Calibrate and analyse existing baskets based on the following model:
\begin{center}\vspace{-1.5em}
\begin{minipage}[t]{.45\textwidth}
  \begin{align*}
         &Y_{k_0}\sim\text{Binomial}(n_{k_0},p_{k_0}),&k_0&=1,\ldots,K_0\\
    &\theta_{k_0}=\log\Big(\frac{p_{k_0}}{1-p_{k_0}}\Big),\\
    &\theta_{k_0}=\delta_{k_0} M_{1k_0}+(1-\delta_{k_0})M_{2k_0},\\
    &\delta_{k_0}\sim\text{Bernoulli}(\pi_{k_0}),
  \end{align*}
\end{minipage}%
\begin{minipage}[t]{0.45\textwidth}
\begin{align*}
    &M_{1k_0}\sim\text{N}(\mu,\sigma^2),\\
      &\mu\sim\text{N}(-1.39,10^2),\\
    &\sigma\sim \text{Half-Normal}(0,1),\\
     &M_{2k_0}\sim \text{N}(-0.85,4.76),
\end{align*}
    \end{minipage}
    \end{center}
with $\pi_{k_0}=0.5$ for all $k_0=1,\ldots,K_0$. Calibrate and analyse new baskets based on the following model:
\begin{center}\vspace{-1.5em}
\begin{minipage}[t]{.45\textwidth}
  \begin{align*}
         &Y_{k}\sim\text{Binomial}(n_k,p_k),&k&=1,\ldots,K\\
    &\theta_k=\log\Big(\frac{p_k}{1-p_k}\Big),\\
    &\theta_k=\delta_kM_{1k}+(1-\delta_k)M_{2k},\\
    &\delta_k\sim\text{Bernoulli}(\pi_k),
  \end{align*}
\end{minipage}%
\begin{minipage}[t]{0.45\textwidth}
\begin{align*}
    &M_{1k}\sim\text{N}(\mu,\sigma^2),\\
      &\mu\sim\text{N}(-1.39,10^2),\\
    &\sigma\sim \text{Half-Normal}(0,1),\\
     &M_{2k}\sim \text{N}(-0.85,4.76),
\end{align*}
    \end{minipage}
    \end{center}
with $\pi_k=0.5$ for all $k=1,\ldots,K$.\\\\

\section{RCaP: Robust Calibration Procedure for Type I Error Control}
\begin{algorithm}[ht]
\begin{algorithmic}
\caption{RCaP - Calibrate $\Delta_k$ across several simulation scenarios for type I error control.}\label{alg:cal2}
\State \textbf{Data:} Total number of simulation scenarios, $M$, scenarios $\bm{p_1},\ldots,\bm{p_M}$, basket sample sizes $\bm{n}$, number of simulation runs for each scenario, $R$, and integer weights for the scenarios, $\omega_1,\ldots,\omega_M$ and null response rate, $q_0$;
\State \textbf{Initialization:} $\bm{PP_1},\ldots,\bm{PP_K}$ empty vectors\;
\For {$m = 1$ to $M$}
\For {$r=1$ to $R$}
\State Generate data $\bm{X}\sim \text{Binomial}(\bm{p_m},\bm{n})$\;
\State Fit information borrowing model to obtain posterior densities\;
\State Compute the posterior probability of a type I error $\mathbb{P}(p_{mk}>q_0|\bm{X})$, in basket $k$\;
\For {$k=1$ to $K$}
\If {T($p_{mk}\leq q_0$)}
\For {$j=1$ to $\omega_m$}
\State $\bm{PP_k}=\bm{PP_k}\cup \mathbb{P}(p_{mk}>q_0|\bm{X})$\;
\EndFor
\EndIf
\EndFor
\EndFor
\EndFor
\State $\Delta_k$ = $100(1-\alpha)\%$ quantile of $\bm{PP}_k$ for each basket $k$.\;
\State \Return {Cut-off values $\Delta_k$ for each basket $k$;}
\end{algorithmic}
\end{algorithm}

\section{Fixed Scenario: Calibrating Across Scenarios}
Presented here are the full results for the simulation study presented in in the fixed scenario simulation of the main text, in which $\Delta_{k_0}$ and $\Delta_{k\prime}$ are calibrated across simulation scenarios 1-5 in Table \ref{tab:scenarios} in the main text. 

\begin{table}[H]
\centering
\caption{Operating characteristics for the fixed scenario simulation study in the main text.}\label{tab:FixedScenarioResuts1}
\scalebox{0.7}{
\begin{tabular}[t]{lcccccccccccc}
\toprule
&\multicolumn{5}{c}{\% Reject} & FWER & \%  Correct & \multicolumn{5}{c}{Mean Point Estimate (Standard Deviation)} \\
\midrule
\textbf{Sc 1} & \textbf{0.2} & \textbf{0.2} & \textbf{0.2} & \textbf{0.2} & \textbf{0.2}\\
IND & 6.33 & 6.52 & 6.42 & 6.46 & 9.82 & 29.37 & 70.63 & 0.202 (0.068) & 0.202 (0.068) & 0.202 (0.068) & 0.203 (0.067) & 0.200 (0.106)\\
UNPL & 5.81 & 5.75 & 5.75 & 5.69 & 5.26 & 22.47 & 77.53 & 0.202 (0.065) & 0.202 (0.066) & 0.202 (0.065) & 0.202 (0.064) & 0.204 (0.079)\\
PL1(a) & 5.73 & 5.92 & 5.89 & 5.78 & 5.45 & 22.82 & 77.18 & 0.202 (0.065) & 0.202 (0.066) & 0.202 (0.065) & 0.202 (0.064) & 0.204 (0.079)\\
PL2(a) & 6.48 & 6.37 & 6.33 & 6.41 & 5.45 & 25.05 & 74.95 & 0.292 (0.068) & 0.202 (0.068) & 0.203 (0.068) & 0.203 (0.067) & 0.204 (0.079)\\\\
\hline
\textbf{Sc 2} & \textbf{0.4} & \textbf{0.2} & \textbf{0.2} & \textbf{0.2} & \textbf{0.2}\\
IND & 75.68 & 8.58 & 8.87 & 8.62 & 9.82 & 30.51 & 51.11 & 0.380 (0.096) & 0.208 (0.072) & 0.209 (0.071) & 0.209 (0.070) & 0.200 (0.106)\\
UNPL & 73.73 & 7.83 & 8.12 & 8.01 & 7.07 & 25.47 & 52.69 & 0.376 (0.096) & 0.208 (0.069) & 0.209 (0.069) & 0.209 (0.068) & 0.212 (0.083)\\
PL1(a) & 74.11 & 8.11 & 8.35 & 8.32 & 7.51 & 26.19 & 52.35 & 0.376 (0.096) & 0.208 (0.069) & 0.209 (0.069) & 0.209 (0.068) & 0.212 (0.083)\\
PL2(a) & 75.67 & 8.49 & 8.70 & 8.62 & 7.38 & 27.58 & 52.91 & 0.380 (0.096) & 0.208 (0.072) & 0.209 (0.071) & 0.209 (0.070) & 0.212 (0.083)\\\\
\hline
\textbf{Sc 3} & \textbf{0.4} & \textbf{0.4} & \textbf{0.2} & \textbf{0.2} & \textbf{0.2}\\
IND & 80.95 & 80.17 & 10.31 & 10.18 & 9.82 & 26.15 & 37.97 & 0.386 (0.092) & 0.385 (0.093) & 0.216 (0.075) &	0.216 (0.074) &	0.200 (0.106)\\
UNPL & 79.17&	78.44&	10.46&	10.06&	10.74&	26.14&	44.87&	0.381 (0.092)&	0.380 (0.092)&	0.216 (0.072)&	0.216 (0.072)&	0.221 (0.087)\\
PL1(a) & 79.44&	78.85&	10.66&	10.36&	11.10&	26.82&	44.77&	0.381 (0.092)&	0.380 (0.092)&	0.216 (0.072)&	0.216 (0.072)&	0.221 (0.087)\\
PL2(a) & 80.95&	80.15&	10.34&	10.04&	11.18&	27.06&	46.64&		0.386 (0.092)&	0.385 (0.093)&	0.216 (0.075)&	0.216 (0.074)&	0.221 (0.087)\\\\
\hline
\textbf{Sc 4} & \textbf{0.4} & \textbf{0.4} & \textbf{0.4} & \textbf{0.2} & \textbf{0.2}\\
IND & 82.50&	81.89&	82.74&	14.15&	9.82&	22.64	&41.66&		0.392 (0.088)&	0.391 (0.089)&	0.392 (0.087)&	0.223 (0.078)&	0.200 (0.106)\\
UNPL & 82.34	&81.77&	82.57	&13.62	&12.57&	23.99&	40.52	&	0.387 (0.087)&	0.386 (0.088)&	0.387 (0.087)&	0.224 (0.076)	&0.231 (0.092)\\
PL1(a) & 82.86	&82.13	&82.78	&13.88	&12.68&	24.32	&40.75	&	0.387 (0.087)&	0.386 (0.088)	&0.387 (0.087)	&0.224 (0.076)	&0.231 (0.092)\\
PL2(a) & 82.68	&82.04	&82.66&	14.24&	12.69&	25.18&	40.15&		0.392 (0.088)	&0.391 (0.088)	&0.392 (0.087)&	0.224 (0.078)	&0.231 (0.092)\\\\
\hline
\textbf{Sc 5} & \textbf{0.4} & \textbf{0.4} & \textbf{0.4} & \textbf{0.4} & \textbf{0.2}\\
IND & 86.74	&86.06	&86.86	&86.85	&9.82	&9.82&	54.18	&	0.399 (0.083)&	0.398 (0.084)	&0.399 (0.083)	&0.399 (0.082)&	0.200 (0.106)\\
UNPL & 85.90&	85.35&	85.83	&85.88&	12.82&	12.82&	49.08	&	0.394 (0.082)&	0.393 (0.083)&	0.394 (0.082)&	0.394 (0.081)&	0.241 (0.096)\\
PL1(a) & 86.45	&85.92&	86.12	&86.42	&13.00	&13.00	&50.33&		0.394 (0.082)&	0.393 (0.083)	&0.394 (0.082)&	0.394 (0.081)&	0.241 (0.096)\\
PL2(a) & 86.84&	86.02	&86.56&	86.73&	13.17&	13.17	&51.89	&	0.399 (0.083)&	0.398 (0.084)&	0.399 (0.083)&	0.399 (0.082)&	0.241 (0.096)\\\\
\textbf{Sc 6} & \textbf{0.4} & \textbf{0.4} & \textbf{0.4} & \textbf{0.4} & \textbf{0.4}\\
IND & 86.74	&86.06	&86.86	&86.85&	65.03&		39.58&	&	0.399 (0.083)	&0.398 (0.084)	&0.399 (0.083)&	0.399 (0.082)&	0.400 (0.131)\\
UNPL & 88.57&	88.12&	88.53&	88.51	&72.25	&	47.45&	&	0.399 (0.080)&	0.399 (0.080)	&0.399 (0.080)	&0.400 (0.078)&	0.398 (0.098)\\
PL1(a) & 88.71&	88.41&	88.97	&88.99&	72.52&		48.03&	&	0.399 (0.080)&	0.398 (0.080)	&0.399 (0.079)&	0.399 (0.078)&	0.398 (0.098)\\
PL2(a) & 86.84&	86.02&	86.56	&86.73	&72.46	&	44.33	&&	0.399 (0.083)&	0.398 (0.084)	&0.399 (0.083)&	0.399 (0.082)&	0.398 (0.098)\\\\
\hline
\textbf{Sc 7} & \textbf{0.2} & \textbf{0.2} & \textbf{0.2} & \textbf{0.2} & \textbf{0.4}\\
IND & 6.33&	6.52&	6.42	&6.46	&65.03	&21.49&	50.86	&	0.202 (0.068)&	0.202 (0.068)&	0.202 (0.068)&	0.203 (0.067)&	0.400 (0.131)\\
UNPL & 7.16&	7.28&	7.51&	7.31&	53.41&	24.19&	38.22&		0.207 (0.067)&	0.206 (0.067)&	0.207 (0.067)&	0.207 (0.066)&	0.365 (0.119)\\
PL1(a) & 7.48&	7.42&	7.59	&7.47&	53.88	&24.67&	38.13	&	0.207 (0.067)	&0.206 (0.067)	&0.207 (0.067)	&0.207 (0.066)	&0.365 (0.119)\\
PL2(a) & 6.48&	6.37&	6.33	&6.41&	53.84&	21.29&	40.94&		0.202 (0.068)&	0.202 (0.068)&	0.203 (0.068)&	0.203 (0.067)&	0.365 (0.119)\\\\
\hline
\textbf{Sc 8} & \textbf{0.4} & \textbf{0.2} & \textbf{0.2} & \textbf{0.2} & \textbf{0.4}\\
IND & 75.68&	8.58&	8.87&	8.62	&65.03&	22.89&	36.84	&	0.380 (0.096)&	0.208 (0.072)&	0.209 (0.071)&	0.209 (0.070)	&0.400 (0.131)\\
UNPL & 77.61&	9.43&	9.53&	9.61&	58.07&	23.97	&32.17		&0.379 (0.093)	&0.213 (0.071)&	0.214 (0.071)&	0.214 (0.070)&	0.372 (0.115)\\
PL1(a) & 77.73	&9.75&	9.75	&9.75&	59.16&	24.28&	33.31	&	0.379 (0.093)&	0.213 (0.071)	&0.214 (0.071)	&0.214 (0.070)	&0.372 (0.115)\\
PL2(a) & 75.67	&8.49	&8.70	&8.62&	59.00&	22.81&	32.46	&	0.380 (0.096)&	0.208 (0.072)	&0.209 (0.071)&	0.209 (0.070)&	0.372 (0.115)\\\\
\bottomrule
\end{tabular}}
\end{table}%

\newpage
\begin{table}[H]
\centering
\caption{Operating characteristics for the fixed scenario simulation study in the main text.}\label{tab:FixedScenarioResuts2}
\scalebox{0.7}{
\begin{tabular}[t]{lcccccccccccc}
\toprule
&\multicolumn{5}{c}{\% Reject} & FWER & \%  Correct & \multicolumn{5}{c}{Mean Point Estimate (Standard Deviation)} \\
\midrule

\textbf{Sc 9} & \textbf{0.4} & \textbf{0.4} & \textbf{0.2} & \textbf{0.2} & \textbf{0.4}\\
IND & 80.91	&80.19	&10.3&	10.12	&64.80&	18.17	&34.09	&	0.385 (0.092)	&0.385 (0.093)	&0.216 (0.075)&	0.216 (0.074)&	0.399 (0.131)\\
UNPL & 81.61&	80.89&	12.17	&12.00 &	67.04&	21.26&	34.35	&	0.385 (0.089)&	0.384 (0.089)&	0.222 (0.075)&	0.221 (0.074)&	0.381 (0.109)\\
PL1(a) & 81.86&	81.04&	12.78&	12.3&	67.93&	22.05&	34.41	&	0.385 (0.089)&	0.384 (0.089)	&0.222 (0.075)&	0.222 (0.074)	&0.381 (0.109)\\
PL2(a) & 80.95	&80.15&	10.34&	10.04&	67.65&	18.14&	36.79		&0.386 (0.092)&	0.385 (0.093)	&0.216 (0.075)&	0.216 (0.074)&	0.381 (0.109)\\\\
\hline
\textbf{Sc 10} & \textbf{0.4} & \textbf{0.4} & \textbf{0.4} & \textbf{0.2} & \textbf{0.4}\\
IND & 82.68	&82.04&	82.78&	14.37&	64.80&	14.37&	30.36	&	0.392 (0.088)&	0.391 (0.089)&	0.392 (0.087)&	0.223 (0.078)&	0.399 (0.131)\\
UNPL & 84.43	&83.79&	84.46&	16.33	&71.47	&16.33&	36.58&		0.393 (0.084)&	0.392 (0.085)	&0.392 (0.084)&	0.228 (0.078)&	0.390 (0.104)\\
PL1(a) & 84.83&	84.30&	84.90	&16.67&	71.77&	16.67&	37.28&		0.393 (0.084)	&0.392 (0.085)	&0.392 (0.084)	&0.228 (0.078)	&0.390 (0.104)\\
PL2(a) & 82.68&	82.04&	82.66&	14.24&	71.77	&14.24	&33.97&		0.392 (0.088)&	0.391 (0.088)	&0.392 (0.087)	&0.224 (0.078)	&0.390 (0.104)\\\\
\hline
\textbf{Sc 11} & \textbf{0.3} & \textbf{0.2} & \textbf{0.2} & \textbf{0.2} & \textbf{0.2}\\
IND & 36.38	&7.73	&7.98	&7.84	&9.83&	28.40&	23.71	&	0.287 (0.083)	&0.206 (0.069)	&0.207 (0.069)	&0.207 (0.068)&	0.200 (0.106)\\
UNPL & 34.33&	7.12&	7.19&	7.33	&6.32&	22.86&	23.15	&	0.284 (0.081)&	0.206 (0.067)	&0.207 (0.066)	&0.207 (0.065)&	0.210 (0.080)\\
PL1(a) & 34.77&	7.29	&7.38	&7.39&	6.63&	23.42&	23.22	&	0.284 (0.081)&	0.206 (0.067)&	0.207 (0.066)&	0.207 (0.065)	&0.210 (0.080)\\
PL2(a) & 36.32	&7.74&	7.88	&7.73&	6.51	&24.98&	24.33	&	0.287 (0.083)&	0.206 (0.069)	&0.207 (0.069)&	0.207 (0.068)&	0.210 (0.080)\\\\
\hline
\textbf{Sc 12} & \textbf{0.3} & \textbf{0.3} & \textbf{0.2} & \textbf{0.2} & \textbf{0.2}\\
IND & 40.65	&40.61&	9.20&	8.87	&9.83	&24.55&	13.63	&	0.291 (0.081)&	0.291 (0.082)	&0.212 (0.070)	&0.212 (0.069)&	0.200 (0.106)\\
UNPL & 38.59	&38.36	&8.78	&8.56	&7.85	&21.50	&12.14	&	0.288 (0.079)	&0.287 (0.080)	&0.212 (0.067)&	0.212 (0.067)&	0.215 (0.082)\\
PL1(a) & 39.15&	38.87	&8.95&	8.78	&8.38	&22.16	&12.23	&	0.288 (0.079)&	0.287 (0.080)	&0.212 (0.067)&	0.212 (0.067)&	0.215 (0.081)\\
PL2(a) & 40.72&	40.36&	9.16&	8.94&	8.34&	22.83&	13.34	&	0.291 (0.081)	&0.290 (0.082)&	0.212 (0.070)&	0.212 (0.069)	&0.215 (0.081)\\\\
\hline
\textbf{Sc 13} & \textbf{0.3} & \textbf{0.2} & \textbf{0.2} & \textbf{0.2} & \textbf{0.3}\\
IND & 36.38	&7.73&	7.98	&7.84	&34.58&	20.65&	8.66&		0.287 (0.083)	&0.206 (0.069)	&0.207 (0.069)&	0.207 (0.068)	&0.300 (0.122)\\
UNPL & 36.70&	8.00&	8.26	&8.05&	26.43	&20.68	&7.47	&	0.287 (0.080)&	0.209 (0.067)&	0.210 (0.067)	&0.210 (0.066)&	0.285 (0.097)\\
PL1(a) & 37.36&	8.37&	8.53	&8.38&	27.08&	21.40&	7.77	&	0.287 (0.080)&	0.209 (0.067)	&0.210 (0.067)&	0.210 (0.066)&	0.285 (0.097)\\
PL2(a) & 36.32	&7.74&	7.88&	7.73	&27.16&	20.54&	7.18	&	0.287 (0.083)	&0.206 (0.069)	&0.207 (0.069)	&0.207 (0.068)	&0.285 (0.097)\\\\
\hline
\textbf{Sc 14} & \textbf{0.3} & \textbf{0.3} & \textbf{0.2} & \textbf{0.2} & \textbf{0.3}\\
IND & 40.65&	40.61&	9.20&	8.87&	34.58&	16.34	&5.08	&	0.291 (0.081)&	0.291 (0.082)&	0.212 (0.070)&	0.212 (0.069)&	0.300 (0.122)\\
UNPL & 40.97&	40.71&	9.93	&9.58	&30.72&	17.53&	5.32&		0.291 (0.078)&	0.290 (0.079)	&0.215 (0.068)&	0.215 (0.067)	&0.290 (0.095)\\
PL1(a) & 41.60	&41.16	&10.33&	9.82&	31.45&	17.96&	5.62&		0.291 (0.078)	&0.290 (0.079)	&0.215 (0.068)&	0.215 (0.067)&	0.290 (0.095)\\
PL2(a) & 40.72&	40.36&	9.16&	8.94&	31.36&	16.37&	5.79	&	0.291 (0.081)&	0.290 (0.082)	&0.212 (0.070)&	0.212 (0.069)&	0.290 (0.095)\\\\
\hline
\textbf{Sc 15} & \textbf{0.4} & \textbf{0.3} & \textbf{0.2} & \textbf{0.2} & \textbf{0.3}\\
IND & 78.98&	42.68&	9.82&	9.60&	34.58&	17.44&	9.32	&	0.382 (0.093)	&0.295 (0.083)&	0.214 (0.072)&	0.214 (0.072)&	0.300 (0.122)\\
UNPL & 78.74&	42.73&	10.77&	10.45	&33.14&	18.92&	9.94&		0.378 (0.091)&	0.295 (0.080)	&0.217 (0.071)&	0.217 (0.070)&	0.295 (0.096)\\
PL1(a) & 79.17	&43.29&	11.11&	10.86	&33.89	&19.54&	10.03&		0.378 (0.091)&	0.295 (0.080)	&0.217 (0.071)&	0.217 (0.070)&	0.295 (0.096)\\
PL2(a) & 78.84&	42.72&	9.79&	9.61&	34.00&	17.46	&10.98	&	0.382 (0.093)&	0.295 (0.083)	&0.214 (0.072)	&0.214 (0.071)&	0.295 (0.096)\\\\
\hline
\textbf{Sc 16} & \textbf{0.4} & \textbf{0.3} & \textbf{0.3} & \textbf{0.2} & \textbf{0.3}\\
IND & 81.03	&44.59	&44.79	&11.75	&34.58&	11.75&	4.65	&	0.383 (0.090)&	0.299 (0.081)&	0.300 (0.080)&	0.219 (0.073)&	0.300 (0.122)\\
UNPL & 81.36&	45.96&	46.39&	13.08&	37.22	&13.08&	7.89	&	0.390 (0.088)&	0.299 (0.078)&	0.300 (0.077)&	0.222 (0.071)	&0.301 (0.094)\\
PL1(a) &81.78&	46.94&	47.1	&13.46&	37.88	&13.46&	8.55&		0.380 (0.088)&	0.300 (0.078)&	0.300 (0.077)&	0.222 (0.071)&	0.301 (0.094) \\
PL2(a) & 80.93	&44.52	&44.87&	11.83&	37.84&	11.83&	5.69&		0.383 (0.091)&	0.299 (0.081)&	0.300 (0.080)&	0.219 (0.073)&	0.301 (0.094)\\\\
\bottomrule
\end{tabular}}
\end{table}%

\section{Comparison of Calibration Across Scenarios 1-5 to 1-10}
Simulation studies in the main text were conducted under the novel robust calibration procedure (RCaP) in order to achieve a 10\% type I error rate on average across scenarios 1-5 (all global and partial nulls assuming equal sample sizes). However, due to the new basket having a reduced sample size, these scenarios no longer cover all partial and global nulls. This is resolved by also including scenarios 7-10 in the calibration procedure. Exploration is now conducted into differences in performance based on the number of scenarios incorporated to average across. 

Note that under UNPL, calibration differs as it consists of just the four existing baskets. The equal sample size across baskets, results in just 4 global and partial null scenarios and thus $\Delta_{k_0}$ is calibrated just across these four scenarios. Results presented incorporate the irrelevant difference between calibration in UNPL, with absolute difference values given as 0 throughout. Cut-off values for new baskets are equal under both calibrations as under scenarios 7-10, it's response rate is effective and thus not included when taking the quantile to obtain $\Delta_{k\prime}$.

\begin{figure}[H]
    \centering 
    \includegraphics[width=0.95\textwidth]{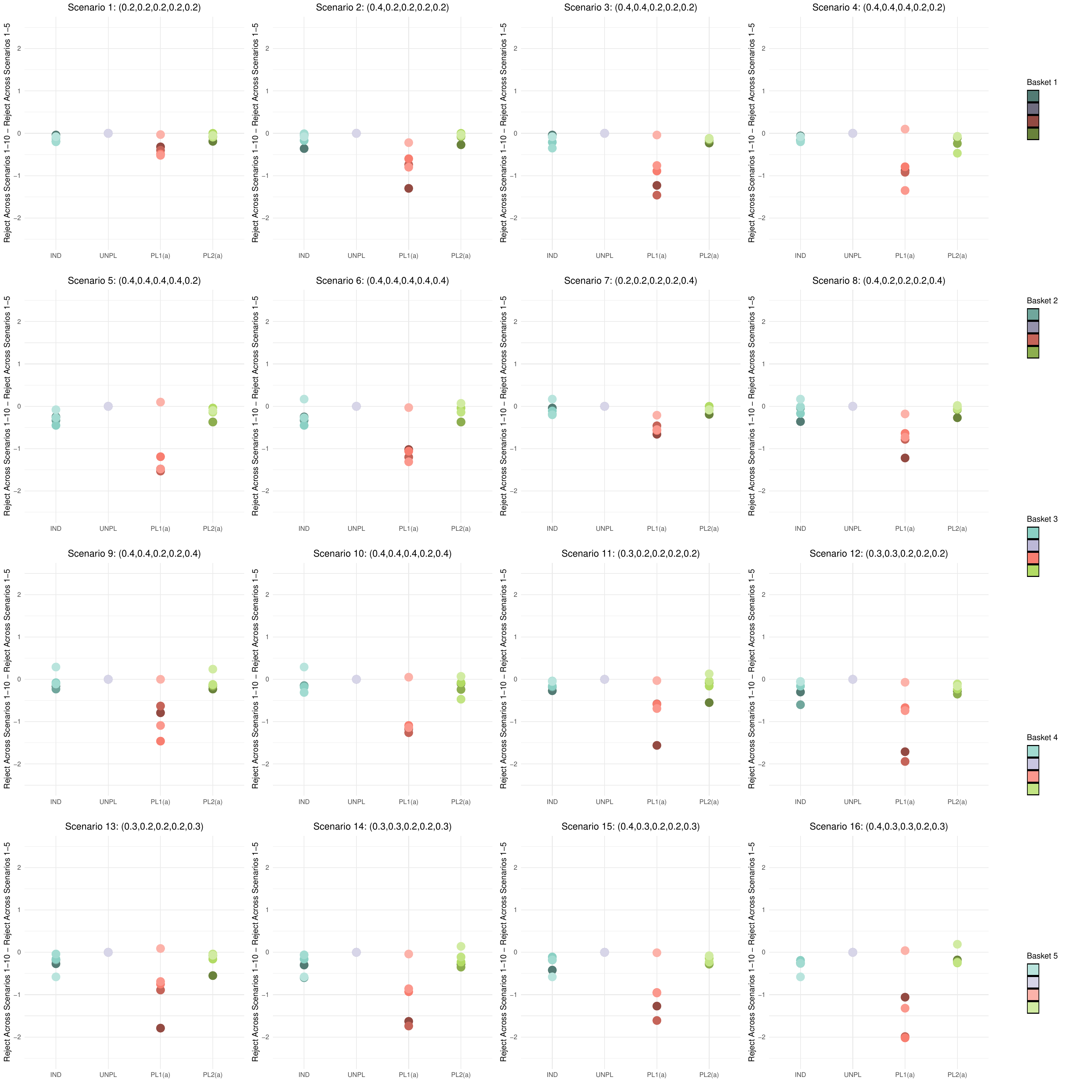}
    \caption{Absolute difference in the number of simulated data sets within which the null hypothesis is rejected between a calibration across scenarios 1-5 and a calibration across scenarios 1-10 (excluding the global alternative). This is split by approach and basket.}
    \label{fig:110v15}
\end{figure}

Figure \ref{fig:110v15} presents the absolute difference in percentage rejections of the null under a calibration across scenarios 1-5 and a calibration across scenarios 1-10 (excluding the global alternative). In all bar a handful of cases, the percentage rejections i.e. type I error rate and power are lower under a calibration across scenarios 1-10 vs a calibration across scenarios 1-5. 

\begin{table}[ht]
    \centering
    \caption{Calibrated $\Delta_{k_0}$ and $\Delta_{k\prime}$ values for each of the approaches for adding a basket under calibration across scenarios 1-5 and calibration across scenarios 1-10.}\label{tab:cut_offsupp}
    \begin{tabular}[t]{lcccc}
    \toprule
         &  \multicolumn{2}{c}{RCaP across 1-10} & \multicolumn{2}{c}{RCaP across 1-5}\\
         & $\Delta_{k_0}$ & $\Delta_{k\prime}$ & $\Delta_{k_0}$ & $\Delta_{k\prime}$\\
         \midrule
         IND & 0.9044 & 0.8989 & 0.9030 & 0.8989\\
         UNPL & 0.9056 & 0.9056 & 0.9056 & 0.9056\\
         PL1(a) & 0.9101 & 0.9021 & 0.9034 & 0.9021\\
         PL2(a) & 0.9044 & 0.9021 & 0.9030 & 0.9021\\
         \bottomrule
    \end{tabular}
\end{table}

Differences under IND and PL2 are always less than 1\%, i.e. a negligible difference. This is expected due to the very similar $\Delta_{k_0}$ and $\Delta_{k\prime}$ values obtained under both calibration cases (see Table \ref{tab:cut_off}). However, more differences are observed under PL1(a), reaching up to 2\% (scenario 16 in a marginally effective basket). This is due to the more conservative cut-off value. So even in the worst cases, differences between approaches are rather small. Calibration across scenarios 1-5 is less computationally expensive as it considers four fewer data scenarios compared to a calibration across scenarios 1-10 (excluding scenario 6 - the global alternative). Due to the very minute differences between approaches, particularly in IND and PL2(a), a calibration across scenarios 1-5 is recommended for it's reduced computational time. 

\newpage
\section{Fixed Scenario: Calibrating Under the Null}
Results presented in the main text utilized the RCaP in which the type I error rate is controlled on average across several data scenarios. The results under a traditional calibration approach in which type I error rate is controlled under a global null scenario, are presented here.

\begin{figure}[H]
    \centering 
    \includegraphics[width=0.95\textwidth]{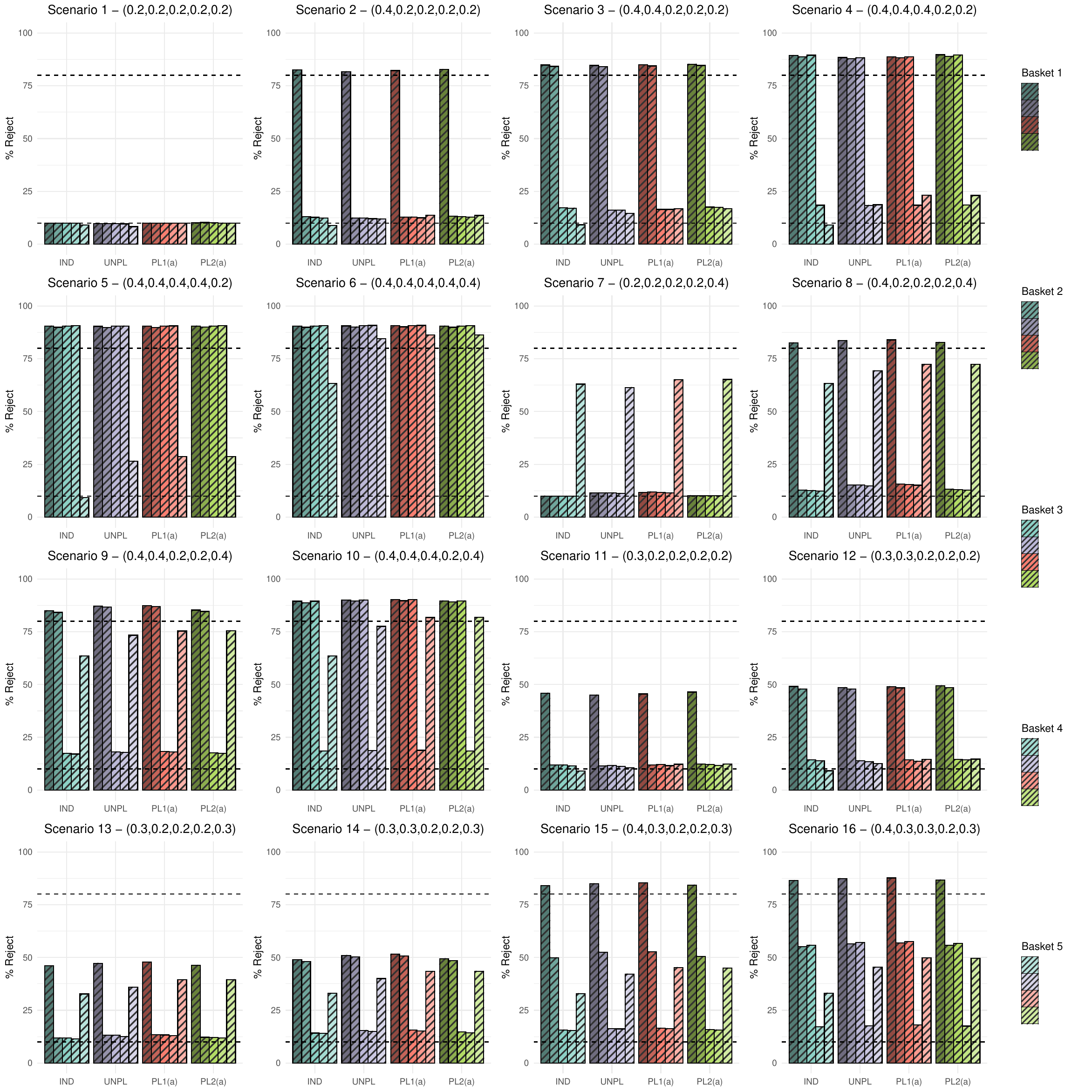}
    \caption{Fixed scenario simulation study results: The percentage of data sets within which the null hypothesis was rejected, where $\Delta$ was calibrated under the null to achieve a 10\% type I error rate on average. This is plotted for each of the four approaches for adding a basket for all five baskets.}
    \label{fig:FixedScenariosNull}
\end{figure}

\newpage
\section{Random Truth Simulation}
\begin{figure}[H]
    \centering 
    \includegraphics[width=1\textwidth]{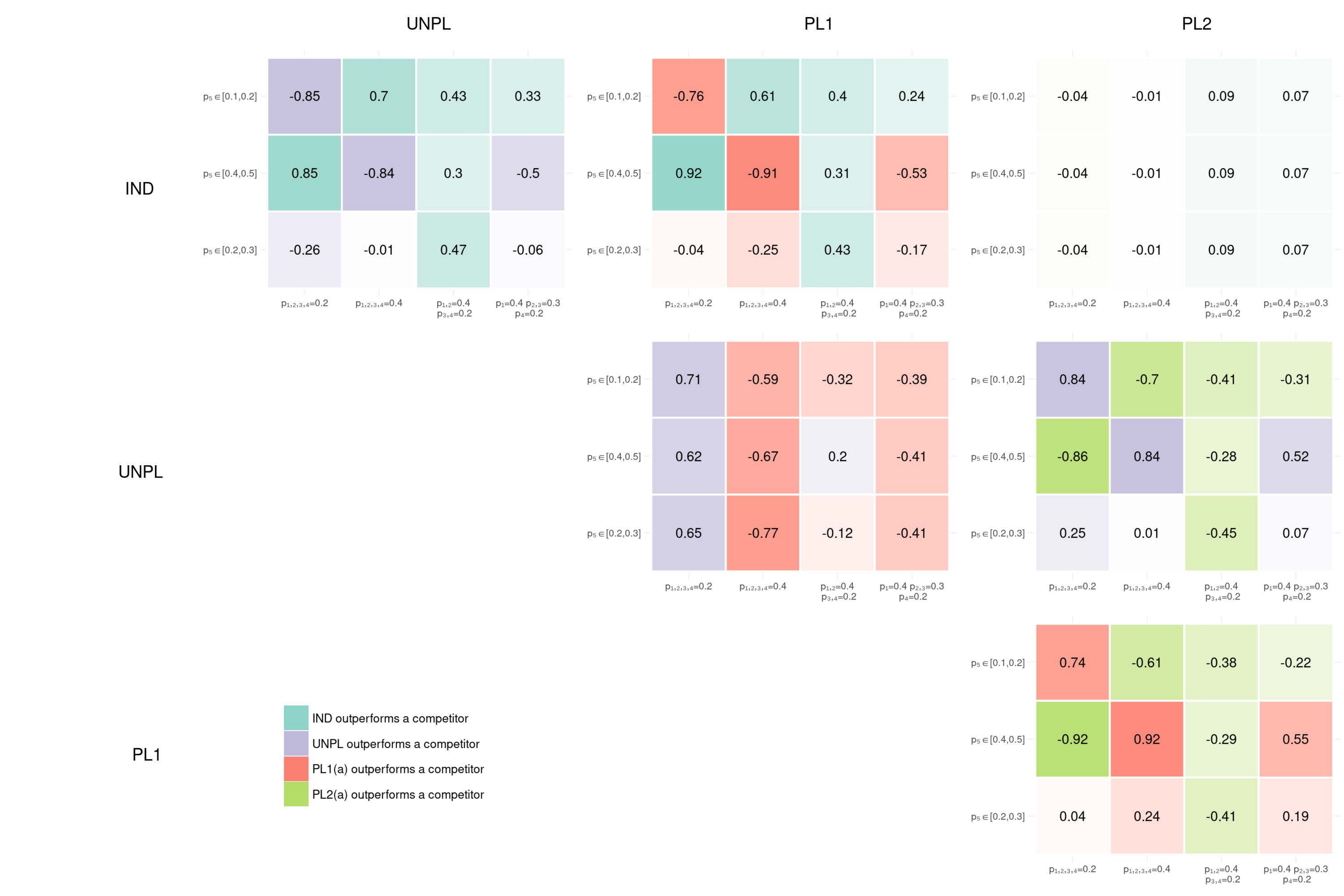}
    \caption{Pair-wise comparison between approaches in each of the 12 simulation settings within which the true response rate in the new basket is varied. The heat map presents the difference in proportion of times the approach corresponding to row gave a correct conclusion over the approach corresponding to column when discrepancies between the two approaches arise in existing baskets only.}
    \label{fig:HeatMapVariedTruthsExisting}
\end{figure}

\begin{figure}[H]
    \centering 
    \includegraphics[width=1\textwidth]{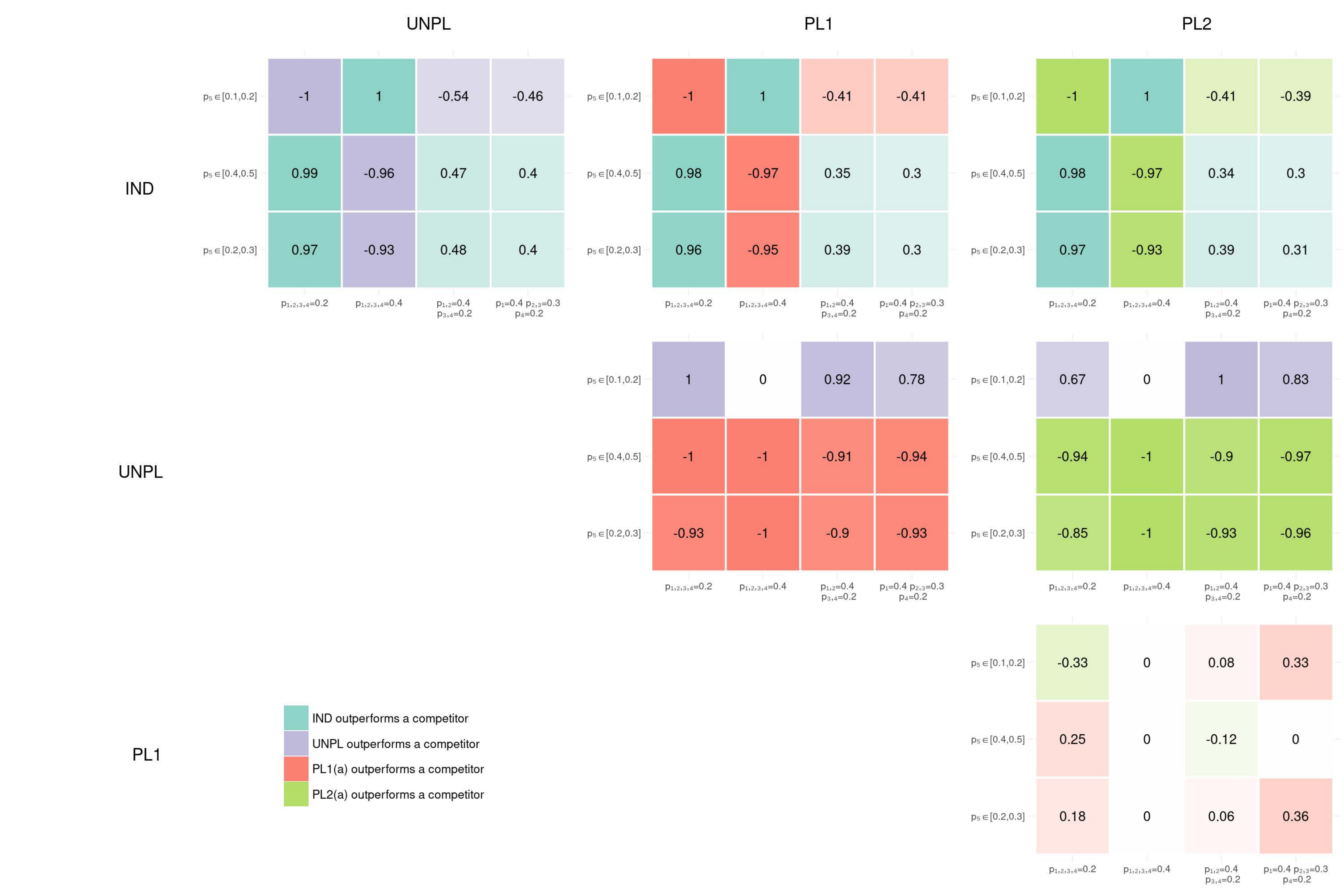}
    \caption{Pair-wise comparison between approaches in each of the 12 simulation settings within which the true response rate in the new basket is varied. The heat map presents the difference in proportion of times the approach corresponding to row gave a correct conclusion over the approach corresponding to column when discrepancies between the two approaches arise in the new basket only.}
    \label{fig:HeatMapVariedTruthsNew}
\end{figure}

Results of pair-wise comparisons between approaches for the simulation study presented in the random truth simulation section of the main text. Figures \ref{fig:HeatMapVariedTruthsExisting} and \ref{fig:HeatMapVariedTruthsNew} present these pair-wise comparisons split into existing and new baskets respectively. Table \ref{tab:VariedScenarioResuts} displays full results for all 12 simulation study settings. 

First consider just existing baskets and pair-wise comparisons. Note that IND and PL2 are equivalent in these baskets, hence values in such a comparison are centered around 0 with slight simulation error. Results for other comparisons are akin to those presented in the main text, indicating that the driving force behind the results presented in the paper are the difference in proportion of correct conclusion when discrepancies lie in existing baskets. 

Then looking at pair-wise comparisons in just the singular new basket, in this case PL1 and PL2 are equivalent and so results are centred around 0 but with rather a lot of simulation noise. In the comparison between IND and UNPL, some cases result in all correct conclusions occurring for just one of the two approaches in discrepancies. For example, in the case where homogeneity between new and existing baskets with all having a null response rate, UNPL in which information is borrowed between all baskets leads to correct conclusions in all 309 cases of discrepancies. Whilst in the case of heterogeneity when the existing baskets are effective with the new basket ineffective, IND where the new is analysed independently leads to the correct conclusion in all 40 discrepancies. The number of cases where UNPL outperforms IND differs when looking at just the new basket compared to all discrepancies, with simulations in which the new basket is ineffective now often preferring UNPL. 

Much more substantial differences are observed in the comparison between UNPL and PL1 under just the new basket compared to overall discrepancies. Previously, in all cases bar when the existing baskets are all null, PL1 outperformed UNPL, i.e. a planned addition is preferred to unplanned. However, when considering just the new basket this reverses with UNPL now only preferred when the new basket is ineffective. This arises from the more conservative $\Delta_{k\prime}$ cut-off under UNPL compared to PL1(a). Note that in most cases very few discrepancies between conclusions under both approaches arise. For instance, when all existing baskets are effective no discrepancies arise when the new basket is ineffective and only 1 or 2 discrepancies arising when it is either marginally effective or effective. 

Similar comparisons between UNPL and PL2 can be drawn as between UNPL and PL1 with cases in which UNPL outperforms PL1 also resulting in a conclusions that UNPL outperformed PL2. 

Pair-wise comparisons between IND and PL1 approaches and IND and PLS approaches result in the same conclusions as those made in the main text.

\begin{table}[H]
\centering
\caption{Overall error rates and power for the varied truth simulation study in which the truth in the new basket is varied with the response rate in existing baskets fixed.}\label{tab:VariedScenarioResuts}
\scalebox{0.8}{
\begin{tabular}[t]{lccccccc}
\toprule
&\multicolumn{5}{c}{\% Reject} & FWER & \% All Correct \\
\midrule
\textbf{Setting 1(a)} & \textbf{0.2} & \textbf{0.2} & \textbf{0.2} & \textbf{0.2} & \textbf{[0.2,0.3]}\\
IND & 6.40 & 6.39 & 6.29 & 6.35 & 24.33 & 21.09 & 19.17\\
UNPL & 5.94 & 6.09 & 5.99 & 6.19 & 12.97 & 19.96 & 8.79\\
PL1(a) & 6.16 & 6.39 & 6.23 & 6.44 & 13.23 & 20.83 & 8.76\\
PL2(a) & 6.36 & 6.41 & 6.29 & 6.33 & 13.19 & 21.12 & 9.49\\
\hline
\textbf{Setting 1(b)} & \textbf{0.2} & \textbf{0.2} & \textbf{0.2} & \textbf{0.2} & \textbf{[0.4,0.5]}\\
IND & 6.40 & 6.39 & 6.29 & 6.35 & 80.73 & 21.08 & 63.45\\
UNPL & 7.45 & 7.33 & 7.57 & 7.43 & 66.96 & 24.76 & 48.18\\
PL1(a) & 7.67 & 7.52 & 7.77 & 7.65 & 67.33 & 25.43 & 47.88\\
PL2(a) & 6.36 & 6.41 & 6.29 & 6.33 & 67.29 & 21.12 & 51.67\\
\hline
\textbf{Setting 1(c)} & \textbf{0.2} & \textbf{0.2} & \textbf{0.2} & \textbf{0.2} & \textbf{[0.1,0.2]}\\
IND & 6.40 & 6.39 & 6.29 & 6.35 & 5.11 & 24.95 & 75.05\\
UNPL & 5.45 & 5.34 & 5.29 & 5.33 & 2.02 & 19.05 & 80.95\\
PL1(a) & 5.58 & 5.60 & 5.57 & 5.50 & 2.07 & 19.76 & 80.24\\
PL2(a) & 6.36 & 6.41 & 6.29 & 6.33 & 2.06 & 22.41 & 77.59\\
\hline
\textbf{Setting 2(a)} & \textbf{0.4} & \textbf{0.4} & \textbf{0.4} & \textbf{0.4} & \textbf{[0.2,0.3]}\\
IND & 88.86 & 86.00 & 86.81 & 86.87 & 24.33 & 0.02 & 15.24\\
UNPL & 86.74 & 86.17 & 86.85 & 86.91 & 25.75 & 0.03 & 17.35\\
PL1(a) & 87.24 & 86.82 & 87.44 & 87.54 & 25.76 & 0.03 & 17.51\\
PL2(a) & 86.86 & 86.02 & 86.79 & 86.90 & 25.76 & 0.03 & 16.03\\
\hline
\textbf{Setting 2(b)} & \textbf{0.4} & \textbf{0.4} & \textbf{0.4} & \textbf{0.4} & \textbf{[0.4,0.5]}\\
IND & 86.86 & 86.00 & 86.81 & 86.87 & 80.72 & 0.00 & 49.60\\
UNPL & 88.73 & 88.26 & 88.92 & 88.85 & 82.49 & 0.00 & 53.85\\
PL1(a) & 89.10 & 88.64 & 89.16 & 89.31 & 82.51 & 0.00 & 54.27\\
PL2(a) & 86.86 & 86.02 & 86.79 & 86.90 & 82.51 & 0.00 & 50.58\\
\hline
\textbf{Setting 2(c)} & \textbf{0.4} & \textbf{0.4} & \textbf{0.4} & \textbf{0.4} & \textbf{[0.1,0.2]}\\
IND & 86.86 & 86.00 & 86.81 & 86.87 & 5.11 & 5.11 & 58.09\\
UNPL & 84.50 & 83.80 & 84.59 & 84.65 & 5.51 & 5.51 & 50.19\\
PL1(a) & 84.86 & 84.33 & 85.02 & 85.21 & 5.51 & 5.51 & 51.76\\
PL2(a) & 86.86 & 86.02 & 86.79 & 86.90 & 5.51 & 5.51 & 57.86\\
\hline
\textbf{Setting 3(a)} & \textbf{0.4} & \textbf{0.4} & \textbf{0.2} & \textbf{0.2} & \textbf{[0.2,0.3]}\\
IND & 80.88 & 80.16 & 10.10 & 9.88 & 24.33 & 17.61 & 13.19\\
UNPL & 80.02 & 79.26 & 10.79 & 10.48 & 22.24 & 18.82 & 12.08\\
PL1(a) & 80.34 & 79.75 & 11.08 & 10.76 & 22.81 & 19.29 & 12.30\\
PL2(a) & 80.87 & 80.15 & 10.20 & 9.88 & 22.79 & 17.68 & 13.37\\
\hline
\textbf{Setting 3(b)} & \textbf{0.4} & \textbf{0.4} & \textbf{0.2} & \textbf{0.2} & \textbf{[0.4,0.5]}\\
IND & 80.88 & 80.16 & 10.10 & 9.88 & 80.72 & 17.59 & 43.16\\
UNPL & 81.75 & 81.06 & 11.92 & 11.79 & 78.25 & 20.75 & 40.82\\
PL1(a) & 82.04 & 81.30 & 12.35 & 12.32 & 79.08 & 21.59 & 40.72\\
PL2(a) & 80.87 & 80.15 & 10.20 & 9.88 & 29.15 & 17.66 & 43.58\\
\hline
\textbf{Setting 3(c)} & \textbf{0.4} & \textbf{0.4} & \textbf{0.2} & \textbf{0.2} & \textbf{[0.1,0.2]}\\
IND & 80.88 & 80.16 & 10.10 & 9.88 & 5.11 & 21.81 & 50.96\\
UNPL & 78.84 & 77.95 & 9.62 & 9.25 & 4.47 & 20.55 & 48.73\\
PL1(a) & 79.40 & 78.47 & 9.80 & 9.50 & 4.70 & 21.01 & 49.13\\
PL2(a) & 80.87 & 80.15 & 10.20 & 9.88 & 4.71 & 21.44 & 50.91\\
\hline
\textbf{Setting 4(a)} & \textbf{0.4} & \textbf{0.3} & \textbf{0.3} & \textbf{0.2} & \textbf{[0.2,0.3]}\\
IND & 80. & 86.00 & 86.81 & 86.87 & 24.33 & 11.76 & 3.44\\
UNPL & 80.80 & 44.94 & 44.94 & 12.18 & 22.76 & 12.21 & 4.89\\
PL1(a) & 81.18 & 45.38 & 45.52 & 12.52 & 23.32 & 12.55 & 5.12\\
PL2(a) & 80.77 & 44.42 & 44.57 & 11.89 & 23.22 & 11.92 & 3.69\\
\hline
\textbf{Setting 4(b)} & \textbf{0.4} & \textbf{0.3} & \textbf{0.3} & \textbf{0.2} & \textbf{[0.4,0.5]}\\
IND & 80.91 & 44.39 & 44.46 & 11.74 & 80.72 & 11.74 & 10.95\\
UNPL & 82.42 & 47.06 & 47.63 & 13.90 & 78.88 & 13.90 & 14.40\\
PL1(a) & 82.81 & 47.66 & 48.22 & 14.26 & 79.46 & 14.26 & 15.13\\
PL2(a) & 80.77 & 44.42 & 44.57 & 11.89 & 23.22 & 11.92 & 3.69\\
\hline
\textbf{Setting 4(c)} & \textbf{0.4} & \textbf{0.3} & \textbf{0.3} & \textbf{0.2} & \textbf{[0.1,0.2]}\\
IND & 80.91 & 44.39 & 44.46 & 11.74 & 80.72 & 11.74 & 10.95\\
UNPL & 82.42 & 47.06 & 47.63 & 13.90 & 78.88 & 13.90 & 14.40\\
PL1(a) & 82.81 & 47.66 & 48.22 & 14.26 & 79.46 & 14.26 & 15.13\\
PL2(a) & 80.77 & 44.42 & 44.57 & 11.89 & 79.46 & 11.89 & 11.11\\

\bottomrule
\end{tabular}}
\end{table}%

\newpage
\section{Investigating the Robustness to the Timing of Addition - IND and UNPL approaches}
In the main text, the robustness of PL1 and PL2 to the timing of addition of a new basket was investigated. Presented here is a similar investigation under the IND and UNPL approaches.

Again assume a case with 4 existing and 1 new basket with sample sizes in existing baskets fixed at 24 patients in each. In this setting, the timing of addition is unknown and thus sample sizes are not known during the calibration procedure. A simulation study is now conducted in which results of all plausible sample sizes in the new basket are considered from $n_5=1$ to 24. Calibration of $\Delta_{k_0}$ and $\Delta_{k\prime}$ is conducted for each possible sample size separately.   

\begin{figure}[H]
    \centering 
    \includegraphics[width=1\textwidth]{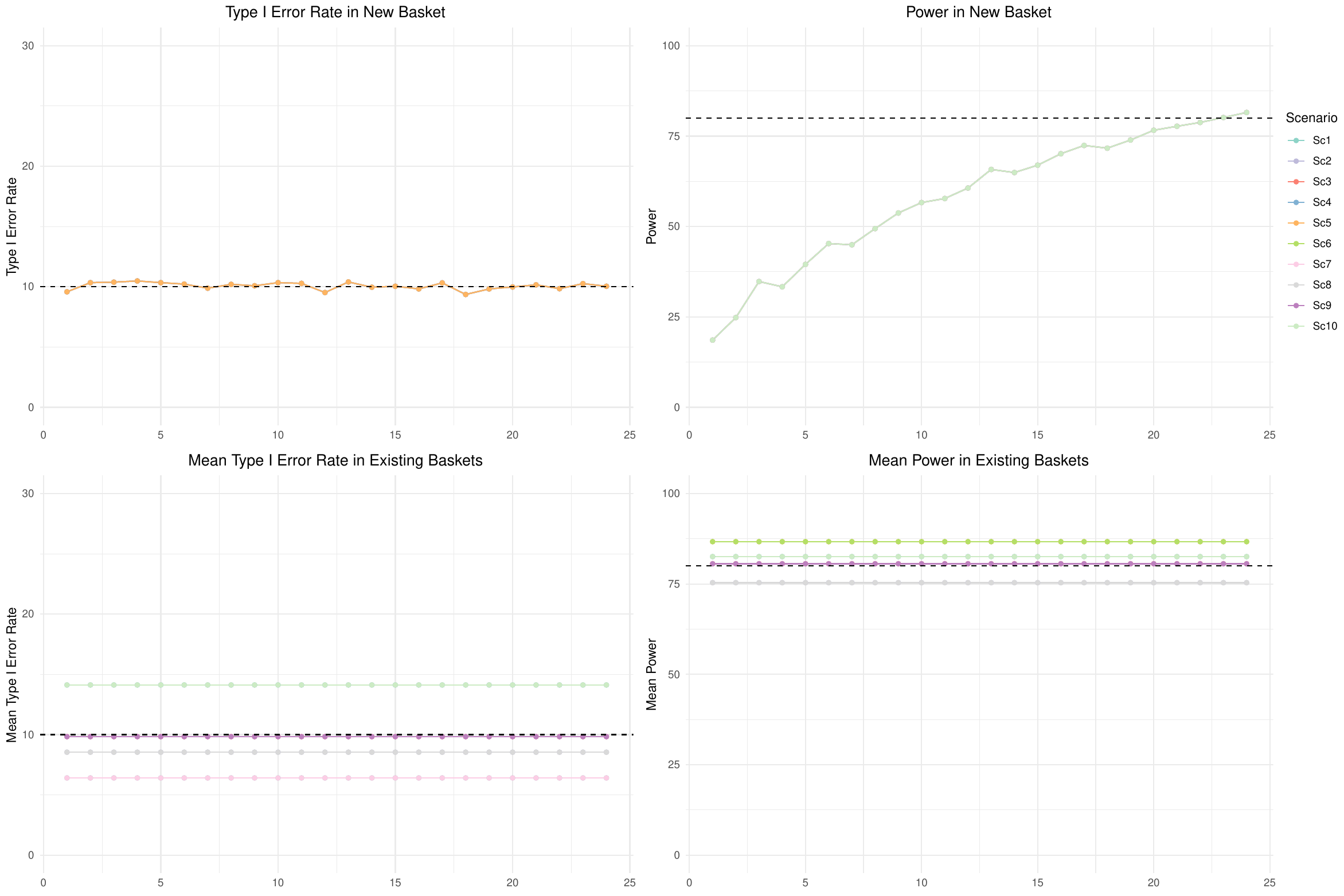}
    \caption{Type I error rate and power under each sample size of $n_5$ from 1 to 24 by applying IND, split by existing and new baskets.}
    \label{fig:Method1SampleSize}
\end{figure}

Figure \ref{fig:Method1SampleSize} presents the change in type I error rate and power as the sample size in the new basket varies, split by new and existing baskets. As the new basket is analyzed independently, the impact of it's sample size on existing baskets is non-existent but also, as each sample size is calibrated to achieve 10\% type I error rate, the impact of change in $n_5$ on error in the new basket is also null. The only variation is in power in the new basket, with larger sample sizes obviously improving power due to the increased certainty in posterior distributions from the added volume of information obtained.

\begin{figure}[H]
    \centering 
    \includegraphics[width=1\textwidth]{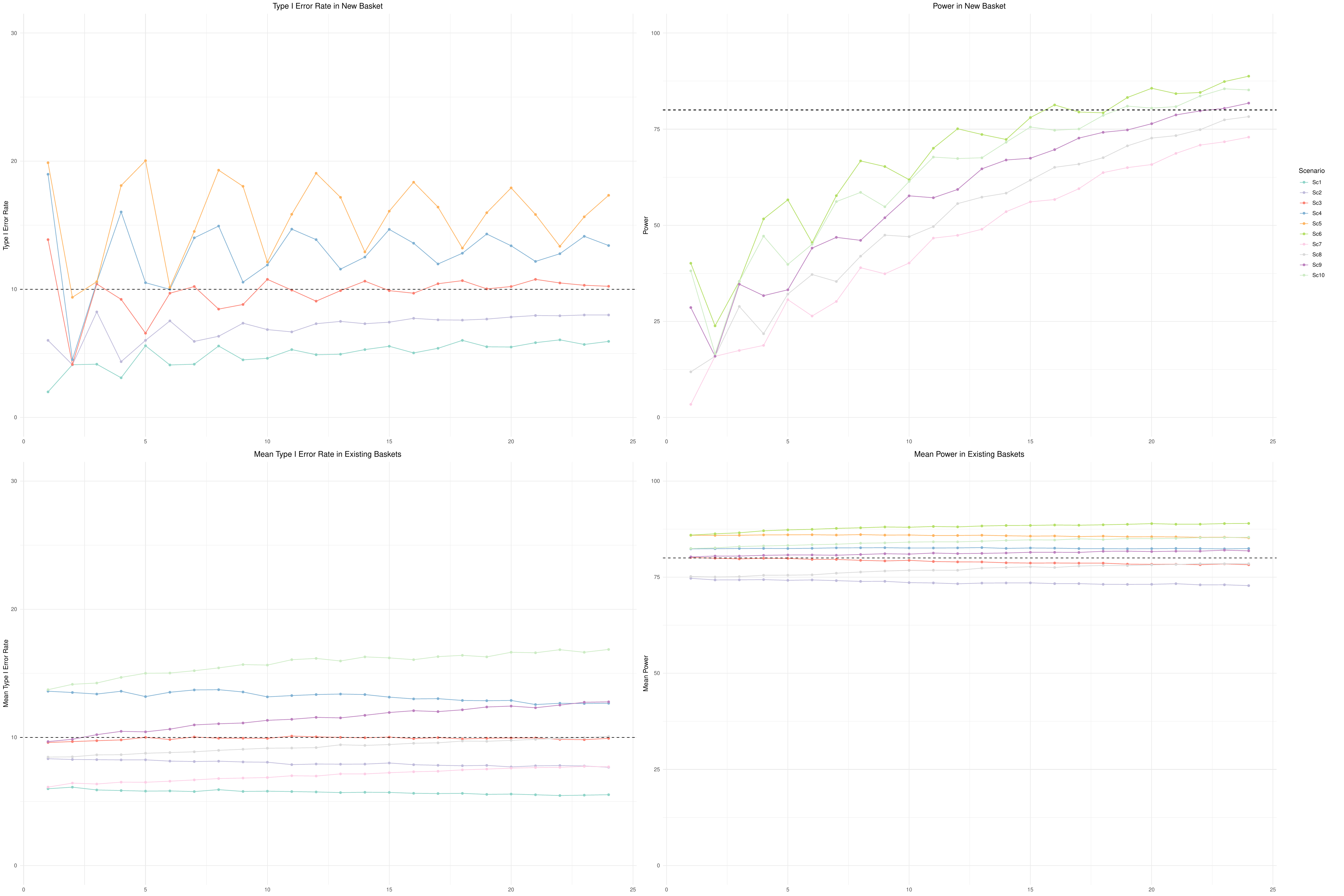}
    \caption{Type I error rate and power under each sample size of $n_5$ from 1 to 24 by applying UNPL, split by existing and new baskets.}
    \label{fig:Method2SampleSize}
\end{figure}

Now under UNPL, the new basket is an unplanned addition and thus the sample size of new baskets has no influence on the calibration procedure. Figure \ref{fig:Method2SampleSize} again presents change in type I error rate and power as $n_5$ varies. Results again imply the sample size in the new baskets has little to no impact on the performance in existing baskets with fairly consistent type I error rates and power across all $n_5$ values. Power in the new basket increases with the sample size as expected and type I error rates form a cyclic pattern due to the discreetness of data. No general increase or decrease in type I error rate is observed as the sample size changes.

\section{Simulation Study - 2 Existing + 2 New Baskets}
All simulation studies conducted so far consisted of four existing baskets opening the trial with one additional basket added during the duration. Instead we now consider a case in which there are two baskets starting the trial with a further two baskets added at a later point. 

\begin{table}[ht]
\centering
\caption{Simulation study scenarios}\label{tab:scenarios22}
\begin{tabular}[t]{lcccc}
\toprule
&$p_1$&$p_2$&$p_3$&$p_4$\\
\midrule
\text{Scenario 1} & 0.2 & 0.2 & 0.2 & 0.2\\
\text{Scenario 2} & 0.4 & 0.2 & 0.2 & 0.2\\
\text{Scenario 3} & 0.4 & 0.4 & 0.2 & 0.2\\
\text{Scenario 4} & 0.4 & 0.4 & 0.4 & 0.2\\
\text{Scenario 5} & 0.2 & 0.2 & 0.4 & 0.2\\
\text{Scenario 6} & 0.4 & 0.2 & 0.4 & 0.2\\
\text{Scenario 7} & 0.2 & 0.2 & 0.4 & 0.4\\
\text{Scenario 8} & 0.4 & 0.2 & 0.4 & 0.4\\
\text{Scenario 9} & 0.4 & 0.4 & 0.4 & 0.4\\
\bottomrule
\end{tabular}
\end{table}%

The same design parameters as previously implemented are used here with a null and target response rate of $q_0=0.2$ and $q_1=0.4$ and a sample size of $n_{k_0}=24$ in existing baskets and $n_{k\prime}=14$ in newly added baskets. Models are specified as outlined in Section \ref{sec:ModelSpec} and data scenarios considered are provided in Table \ref{tab:scenarios22}. Cut-off values $\Delta_{k_0}$ and $\Delta_{k\prime}$ under the IND, PL1 and PL2 approaches, are calibrated across scenarios 1-8 with $\Delta_{k_0}$ taken as the quantile of posterior probabilities across scenarios 1-2 and 5-8 for basket 2 and $\Delta_{k\prime}$ as the quantile across scenarios 1-6 of basket 4. For UNPL, the cut-off value is calibrated across just two scenarios: $p=(0.2,0.2)$ and $p=(0.4,0.2)$.

Now, as multiple baskets are added during the trial, the IND approach gives two options: (a) analyse both new baskets as independent of existing baskets and one another or (b) analyse both new baskets as independent of existing baskets but borrow from each other using a second EXNEX model. 

\begin{figure}[H]
    \centering 
    \includegraphics[width=1\textwidth]{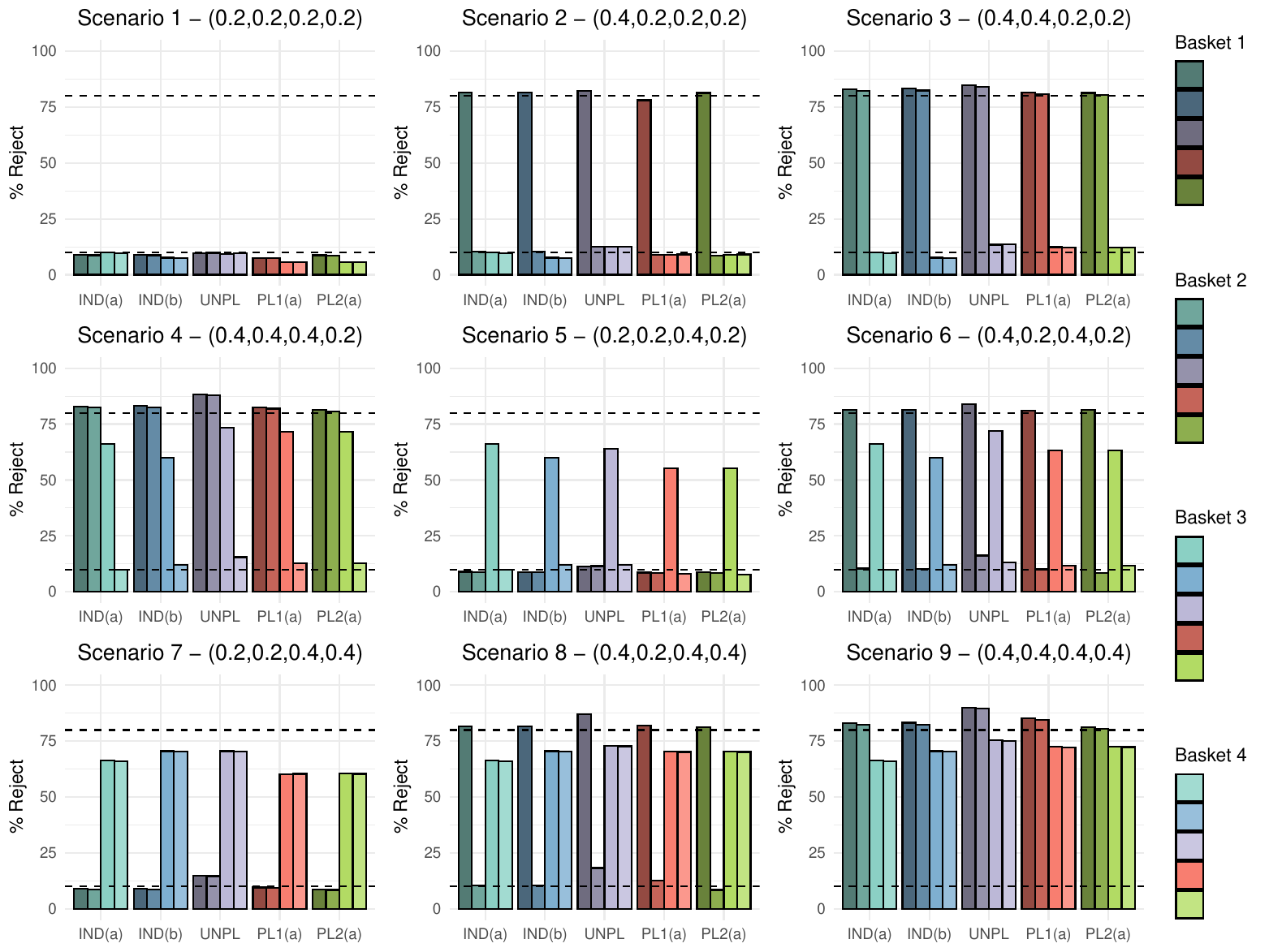}
    \caption{Percentage of data sets within which the null hypothesis was rejected for a simulation study consisting of 2 existing baskets with 2 additional baskets added part-way through the study.}
    \label{fig:2plus2}
\end{figure}

Results for the percentage of data sets within which the null hypothesis was rejected, i.e. type I error rate and power, are presented in Figure \ref{fig:2plus2}. First, consider the differences between IND(a) and IND(b). Under IND(a), due to independent analysis, error rates in the new basket are always controlled to the 10\% level but the same does not hold for IND(b) in which an EXNEX model allows borrowing between both new baskets. In cases where both new baskets are heterogeneous, this leads to reduced error rates lying below the nominal level (e.g. 7.6\% under scenario 1) but in cases of heterogeneity where one new basket is effective and the other ineffective, error rates inflate to approximately 12\%. In these scenarios, power is pulled down from 66.2\% to 60.1\% when utilizing information borrowing. However, significant power can be gained over an independent analysis in cases where both new baskets are effective to treatment (scenarios 7-9). Under IND(a) this power is 66.2\% compared to IND(b) with power 70.6\%. 

Under UNPL, $\Delta_{k_0}=\Delta_{k\prime}=0.865$ compared to 0.900 under PL1(a). This reduced cut-off value leads to less conservative rejections under both new and existing baskets. This results in higher power across all cases, with UNPL giving highest power in all scenarios (e.g. UNPL has a power of 89.8\% and 75.3\% for existing and new baskets respectively under data scenario 9, whereas, PL1(a) has power 84.9\% and 72.4\% respectively). With this, UNPL also possesses the greatest error inflation up to 18.3\%.

Approaches PL1(a) and PL2(a) are equivalent for the new baskets so results differ only in existing baskets. Some cases with more noticeable differences are scenario 2 in which power is increased significantly under PL2 at 81.3\% compared to 78.1\% under PL1 with indistinguishable difference in error; scenario 8 in which both approaches give similar power but PL1 has higher error rates at 12.6\% compared to PL2 at 8.5\% and finally, scenario 9 in which power in existing baskets is greater under PL1 at 84.8\% compared to 80.9\% under PL2.

To conclude, results in this simulation study present fairly similar results to the previous case consisting of 4 existing and one new basket. One of the main differences lies in the UNPL approach which has a far less conservative cut-off than any of the other approaches. This occurs because there are only 2 existing baskets that can be used to calibrate UNPL and as such, estimates lack certainty and only 2 data scenarios are calibrated across. Also displayed in this case, is the potential losses one can make when utilizing IND(b) in all cases bar when both baskets are homogeneous to treatment. 

\end{document}